\documentclass[twocolumn]{aastex631}

\pdfoutput=1 
\usepackage{amsmath,amstext}
\usepackage[T1]{fontenc}
\usepackage[figure,figure*]{hypcap}
\usepackage{xspace}
\usepackage{xcolor}

\newcommand{\rphk}{log($R'_{HK}$)\xspace}
\newcommand{\logg}{log($g$)\xspace}
\newcommand{\teff}{$T_\mathrm{eff}$\xspace}
\newcommand{\feh}{[Fe/H]\xspace}
\newcommand{\vsini}{V$sin(i)$\xspace}
\newcommand{\kms}{km~s^{-1}\xspace}
\newcommand{\cahk}{Ca~II~H~and~K\xspace}
\newcommand{\rstar}{\ensuremath{R_\star}\xspace}

\newcommand{\rsun}{\ensuremath{R_\sun}\xspace}
\newcommand{\msun}{\ensuremath{M_\sun}\xspace}

\newcommand{\rearth}{\ensuremath{R_\mathrm{Earth}}\xspace}

\newcommand{\USQ}{Centre for Astrophysics, University of Southern Queensland, Toowoomba, QLD, Australia}
\newcommand{\CIT}{California Institute of Technology, Pasadena, CA 91125, USA}

\accepted{to the Astrophysical Jouranl, Dec 2023}

\shorttitle{CKS-HK}
\shortauthors{Isaacson et al.}

\begin{document}

\title{The California-Kepler Survey. XI. A Survey of Chromospheric Activity Through the Lens of Precise Stellar Properties}

\author[0000-0002-0531-1073]{Howard Isaacson}
\affiliation{501 Campbell Hall, University of California at Berkeley, Berkeley, CA 94720, USA}
\affiliation{\USQ}
\correspondingauthor{Howard Isaacson}

\email{hisaacson@berkeley.edu}

\author[0000-0002-7084-0529]{Stephen R. Kane}
\affiliation{Department of Earth and Planetary Sciences, University of California, Riverside, CA 92521, USA}
\affiliation{\USQ}

\author[0000-0002-7084-0529]{Brad Carter}
\affiliation{\USQ}

\author[0000-0001-8638-0320]{Andrew W. Howard}
\affiliation{\CIT}

\author[0000-0002-3725-3058]{Lauren Weiss}
\affil{Department of Physics, University of Notre Dame, Notre Dame, IN 46556, USA}

\author[0000-0003-0967-2893]{Erik A. Petigura}
\affiliation{Department of Physics \& Astronomy, University of California Los Angeles, Los Angeles, CA 90095, USA}

\author[0000-0003-3504-5316]{Benjamin Fulton}
\affiliation{NASA Exoplanet Science Institute/Caltech-IPAC, MC 314-6, 1200 E California Blvd, Pasadena, CA 91125, USA}

\begin{abstract}

Surveys of exoplanet host stars are valuable tools for assessing population level trends in exoplanets, and their outputs can include stellar ages, activity, and rotation periods.
We extracted chromospheric activity measurements from the California-Kepler Survey(CKS) Gaia survey spectra in order to probe connections between stellar activity and fundamental stellar properties. Building on the California Kepler Survey’s legacy of 1189 planet host star stellar properties including temperature, surface gravity metallicity and isochronal age, we add measurements of the \cahk lines as a proxy for chromospheric activity for 879 planet hosting stars.  We used these chromospheric activity measurements to derive stellar rotation periods.
We find a discrepancy between photometrically derived and activity-derived rotation periods for stars on the Rossby Ridge. These results support the theory of weakened magnetic braking. We find no evidence for metallicity-dependent activity relations, within the metallicity range of -0.2 to +0.3 dex. With our single epoch spectra we identify stars that are potentially in Maunder Minimum like state using a combination of \rphk and position below the main-sequence. We do not yet have the multi-year time series needed to verify stars in Maunder Minimum like states.
These results can help inform future theoretical studies that explore the relationship between stellar activity, stellar rotation, and magnetic dynamos.

\end{abstract}


\showboxdepth=5
\showboxbreadth=5

\section{Introduction} 
\label{sec:intro}

The study of stellar chromospheres was championed in the late 1960s by the Mt. Wilson S-value project \citep{Vaughan1978, Duncan1991}. With a dedicated telescope for daily observations, the Mt Wilson team began a series of observations that would last decades culminating in the study of stellar activity cycles and comparison to the solar activity cycle. The legacy of the Mt. Wilson HK project was expanded from the photomultiplier era to the era of charge coupled devices (CCDs) by additional studies of stellar activity cycles including those conducted by \citet[][815 stars]{Henry1996}, \citet[][1200 stars]{Wright2004}, \citet[][143 stars]{Hall2007},  \citet[][2630 stars]{Isaacson2010} and more recently \citet[][1674 stars]{GomesdaSilva2021}. These surveys were often secondary to the primary science objectives (\citealt{Hall2007} excepted) of searching for exoplanets  and measuring their masses via high-resolution spectroscopy and precise radial velocities (RVs). For example, the importance of understanding stellar activity as a possible false positive scenario for RV planet mass detection was a primary concern since the first exoplanet discovery \citep{Mayor1995}. In addition to the search for direct correlations between RVs and stellar activity cycles over the timescale of decades, \citep{Wright2008, Fulton2015}, sophisticated analyses such as the FF$\prime$ method, Gaussian processes, and other signal processing algorithms can be employed on data spanning shorter timescales to disentangle the complex relationships that planets have on stars gravitationally from stellar activity due to surface inhomogeneities \citep{Aigrain2012, Howard2013, Pepe2013}.  However, sophisticated signal processing techniques can sometimes overfit the data, and their results should be interpreted with caution \citep{Blunt2023}. When additional information beyond stellar spectra are available, such as space based photometry from \emph{Kepler}, K2 or the Transiting Exoplanet Survey Satellite (TESS) \citep{Kosiarek2020}, ever smaller planets can be characterized with RVs from instruments such as the High Resolution Echelle Spectrometer (HIRES) \citep{Murphy2021} or refuted with instruments such as Habitable Planet Finder \citep{Lubin2021}.
Analysis of stellar flares in photometric data, along with activity metrics from spectra such as H-alpha can be used to study planetary habitability. \citep{Su2022} used H-alpha measurements from LAMOST's low-resolution (1026 stars) and medium-resolution spectra (158 stars)  plus light curves from \emph{Kepler}, K2 and TESS  host stars, and assessed both atmospheric burn off and recovery.

The impact of stellar activity observations has led to greater understanding the relationship of stellar activity, rotation periods, and age. \cite{Noyes1984} laid the foundation for studying connection between activity, convection and rotation, showing that rotation periods correspond to certain levels of activity and both are related to the convective action in solar-type main-sequence stars.  Remarkably, the study used only 40 stars, with what would now be considered primitive determinations for stellar temperature and mass that were based on B-V colors, without parallaxes or high-resolution spectra. Identification of the Rossby number, the ratio of the stellar rotation period to the convective turnover time, was a critical piece of the rotation-activity-age puzzle. This early work, focusing primarily on solar-like, main-sequence stars, was summarized by \cite{Duncan1991} who provided the activity catalog for a large number of stars and \cite{Baliunas1995} who focused on 111 sun-like stars. To extend the age-activity relations beyond solar type stars, \cite{Mamajek2008} determined ages by analyzing young star clusters with ages between 200 Myr and 7 Gyr open clusters using both X-rays and chromospheric activity measurements. As stars age the stellar wind, and more generally magnetic activity, transports angular momentum away from the star, resulting in decreasing rotation periods and lower activity levels. By adding star clusters with various well-known ages, \cite{Mamajek2008} quantified the relationship between stellar spindown and stellar age for stars much younger than the sun. With the use of Gaia proper motion, new young clusters have been identified, adding to the collection of age and activity analyses. \cite{Curtis2020} use open clusters with ages between 0.7 and 1.4 Gyr to identify a pause in the spindown relationships that is especially prominent for lower mass stars. The change in spindown can be accounted for tuning core-envelope models, but other explanations remain possible.

The \emph{Kepler} era of space-based photometric surveys led to the detection of over 4000 transiting planet candidates  \citep{Borucki2011,Borucki2011b, Batalha2013,Thompson2018}. While the first exoplanet systems detected by \emph{Kepler} were confirmed by ground based follow-up observations, \cite{Borucki2010,Batalha2011}, it was the large scale survey of $\sim$200,000 stars by \emph{Kepler} that led to the most important results. 

The California-\emph{Kepler} Survey, a magnitude limited spectroscopic survey of 1189 \emph{Kepler} host stars was undertaken with the focus on improving the uncertainty in stellar radius, and the associated planetary radii. Analysis of high-resolution stellar spectroscopy using Local Thermodynamic Equilibrium (LTE) \cite{Valenti2005} is capable of determining fundamental stellar properties and can be combined with stellar evolution models to determine stellar mass and radius to a typical precision of 10\% and as low as 2\% \citep{CKS2,Berger2020a}.  
The uniform, and homogeneous dataset of high-resolution spectra from the Keck I telescope and HIRES instrument \citep{Petigura2017} has allowed for detection of the detailed structure in the radius distribution for planet sizes between one and four \rearth, \cite{Fulton2017}.  
The CKS dataset has allowed for a series of papers including beyond the planet radius gap including detailed analysis of systems with multiple transiting planets \cite{Weiss2018}, refinement on the Minimum Mass Extra-solar Nebular \citep{Dai2020}, and analysis showing that the Kepler field has similar metallicity to the solar neighborhood \citep{Petigura2018}. 

The eventual addition of Gaia parallaxes further refined the stellar properties of the full \emph{Kepler} sample, their planets' radii and more broadly planet occurrence \citep{Hsu2019, Fulton2018}. With precise Gaia distances the dominant source of error on the planet radius, now on average 5\%, becomes the photometry \citep{Petigura2020}, a major transition compared to the first transiting planets measured with \emph{Kepler} data.  Theoretical studies have suggested that the sub-structure in the planet radius distribution is due to photoevaporation that occurs in the first 100 Myr of planet formation \citep{Lopez2013,Owen2013,Chen2016}. Observational studies calculating precise ages reveal that planet radius changes may continue beyond 1 Gyr when considering the entire \emph{Kepler} planet sample \citep{Berger2020b} and well-defined sub-samples \citep{david2022}. The radius gap as also reported observationally by \cite{VanEylen2018} in an analysis of planet hosts studied with asteroseismology.  

The paper is laid out as follows. Section \ref{sec:sval_extract} describes how we derived S-values from the HIRES spectra and section \ref{sec:sample} describes our star and planet sample. Section \ref{sec:act-stellar-props} shows how the planet properties of CKS-Gaia relate to new stellar activity metrics. Section \ref{sec:rotation_periods} explores how the rotation periods determined from \emph{Kepler} photometry relate to activity metrics from this sample. Activity measurements are correlated with fundamental stellar properties in Section \ref{sec:prot_stellar_props}. In Section \ref{sec:ages} we discuss ages derived from \rphk values and we touch on the least active stars in our sample and discuss implications in section \ref{sec:mm_search}. Finally, in section \ref{sec:planets_activity} shows explores activity and our \emph{Kepler} planet sample.

\section{Methods}

\subsection{S-value Extraction and Calibration}
\label{sec:sval_extract}
\begin{figure*}
\includegraphics[width = 2.0\columnwidth]{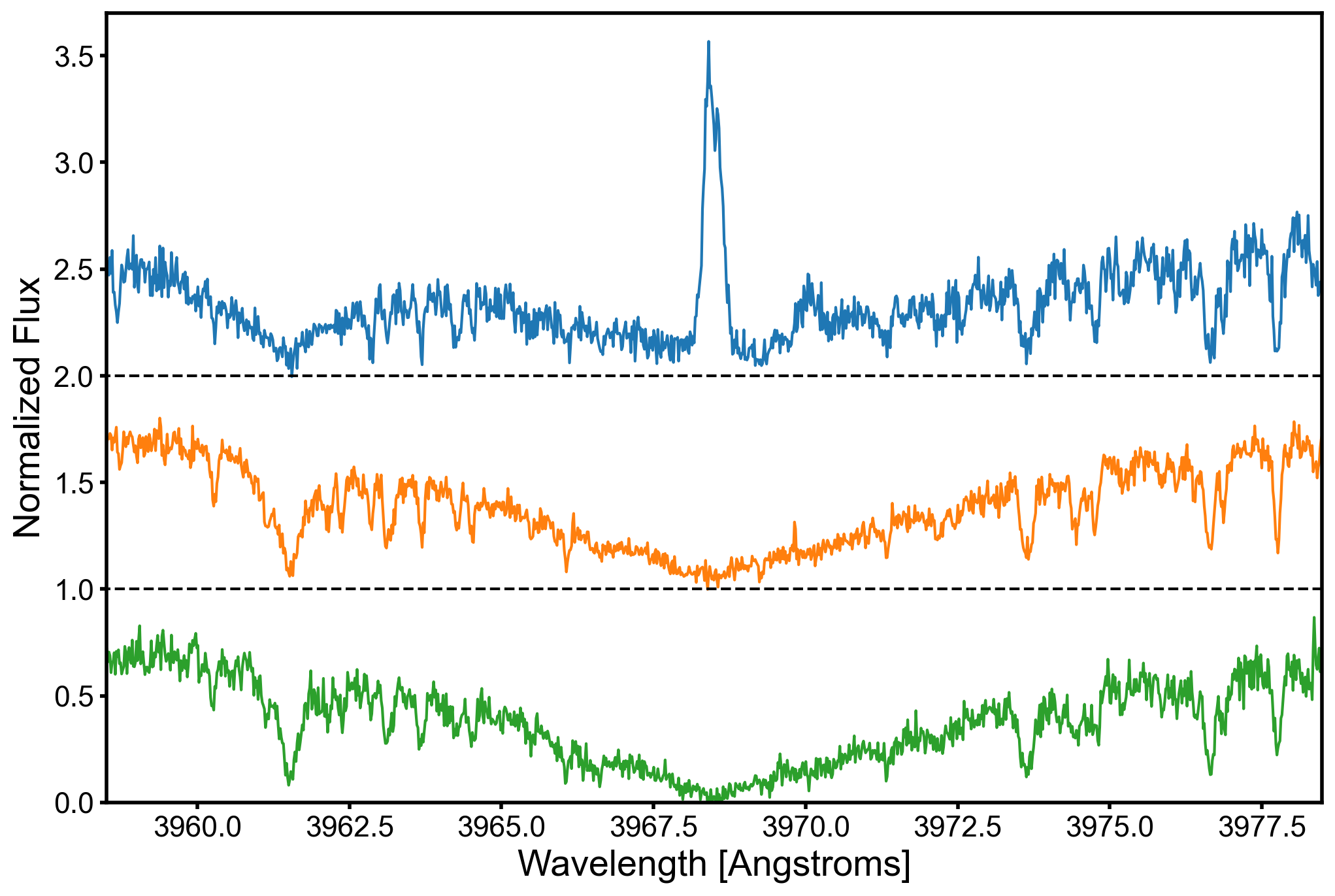}
\caption{ The Ca II H-line is shown for the stars with the highest, median and lowest S-values from top to bottom. They are KOIs-3497, 700, and 629. The SNR of 5-10 per pixel in the continuum sections for these stellar spectra makes the spectral extraction challenging. For active stars, the reversal in the core of the Ca H-line is obvious and rises above the continuum. Small changes in the activity level of low activity stars are challenging to detect due to the lower flux in the line cores.} 
\label{fig:fig_min_med_max_spec}
\end{figure*}

We follow the method of \cite{Isaacson2010} to extract the flux values in the cores of the \cahk lines and continuum regions redward and blueward of the H and K absorption features (Equation \ref{eqn:Sval_basic}, \cite{Vaughan1978}). While this extraction method has been used to analyze post-upgrade HIRES data dating back to 2005, we modified the existing algorithm to optimize the signal to noise ratio (SNR) of single epoch spectra that range from 6 to 10 per reduced pixel. \cite{Isaacson2010} used an SNR cutoff of 5, but only 2\% of spectra were below SNR of 10, so the extraction routine was not well tested for SNR = 5-10. Examples of the \cahk line cores for stars with the highest, median and lowest S-values in our sample are shown in Figure \ref{fig:fig_min_med_max_spec}. This is distinct from the most and least active stars, which is measured by \rphk. Note that low SNR makes the extraction of the fluxes more challenging in spectra with SNR of 5-10 in the continuum sections. In this work, we spline the National Solar Observatory (NSO) solar atlas onto the HIRES rest-frame wavelength solution, and use it as the template to align all other spectra.  The \cite{Isaacson2010} spectral extraction method was optimized for measuring differential S-values of the same star using a high SNR template, the current analysis utilizes the NSO template for all stars in this sample to ensure the absolute scale is as accurate as possible for the single epochs of the CKS-Gaia sample. 

\begin{equation}
\label{eqn:Sval_basic}
S_{HK} = \frac{H + K }{R + V}
\end{equation}

\begin{figure*}
\includegraphics[width = 2.0\columnwidth]{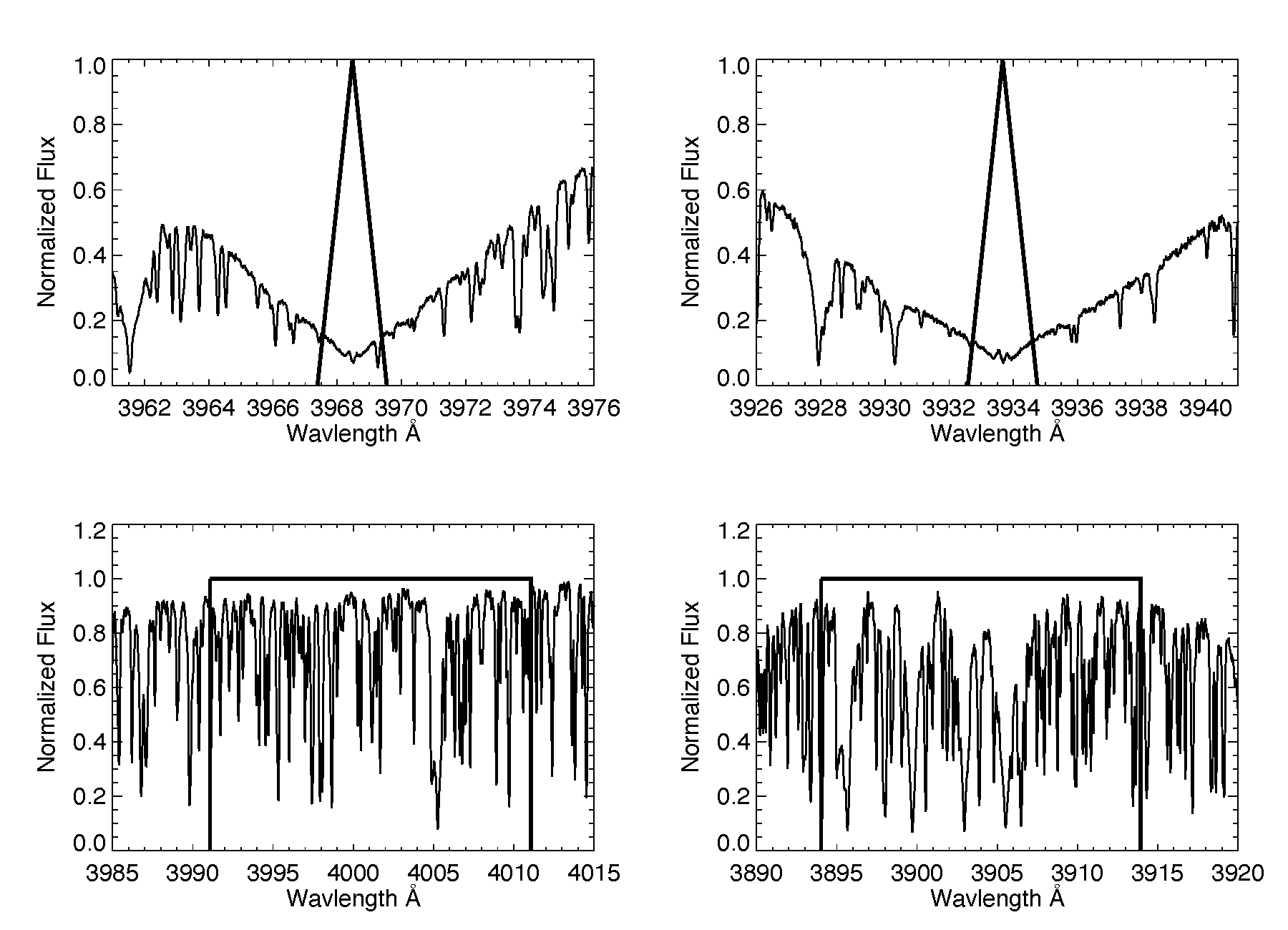}
\caption{ Clockwise from top left: the \cahk lines, and the two continuum sections on either side of the H and K lines, dubbed V (centered at 4000\AA) and R (centered at 3905 \AA) sections. The two 20 \AA ~continuum sections are used to calibrate the variable flux in the 1 \AA~weighted sections in the line cores. Extracting the spectral segments in this way allows for calibration to the Mt. Wilson scale and comparison to other activity surveys.} 
\label{fig:fig_hkrv_4panel}
\end{figure*}

S-values are calculated by summing the flux in the cores of the \cahk lines and dividing by the flux in two continuum sections redward and blueward of the line cores (Figure \ref{fig:fig_hkrv_4panel}). The value of an isolated S-value is difficult to interpret across different spectral types because the intensity of the neighboring regions varies with stellar type, which means the raw S-value index is sensitive to both chromospheric emission and overall SED. So we calculate \rphk,  a metric of chromospheric activity that is comparable across stars with different \teff. \rphk is defined as the base-ten logarithm of the chromospheric portion of the flux in the \cahk line cores relative to the bolometric flux of the star \citep{Noyes1984}. We use the \cahk line flux to measure the non-thermal heating that is related to magnetic activity in the star. By accurately accounting for and subtracting the photospheric contribution to the flux in the cores of the \cahk line cores, we can compare activity across a range of effective temperatures.

With this new flux normalization and extraction method, we require a new calibration to the Mt. Wilson Scale to ensure we can compare our activity metrics on a standard scale. Using a procedure similar to \cite{Isaacson2010}, we use four coefficients and perform a least squares fit for two free parameters, $C_{1}$ and $C_{4}$ in equation \ref{equ:sval_fit1},  with two of the coefficients, $C_{2}$ and $C_{3}$ determined by the ratios of the H to K line fluxes and R to V line fluxes. The final coefficients are shown in Equation \ref{equ:sval_fit1} and were found using 154 stars that were observed on HIRES and Mt. Wilson. We restrict the calibration stars to \teff  between 4700K and 6500K with $ \vsini < 10~\kms $  and a \logg  greater than 4.0, matching the demographics of the majority of the CKS-Gaia sample.

 \begin{equation}
\label{equ:sval_fit1}
S_{HK} = C_{1} * \frac{(H + C_{2}* K)}{(R + C_{3} * V)} + C_{4} 
\end{equation}

\begin{equation}
\label{equ:sval_fit2}
S_{HK} = 22.5 * \frac{(H + 1.01019 * K )}{(R +  1.26134 * V)} - 0.006
\end{equation}

\begin{figure*}
\includegraphics[width = 2.1\columnwidth]{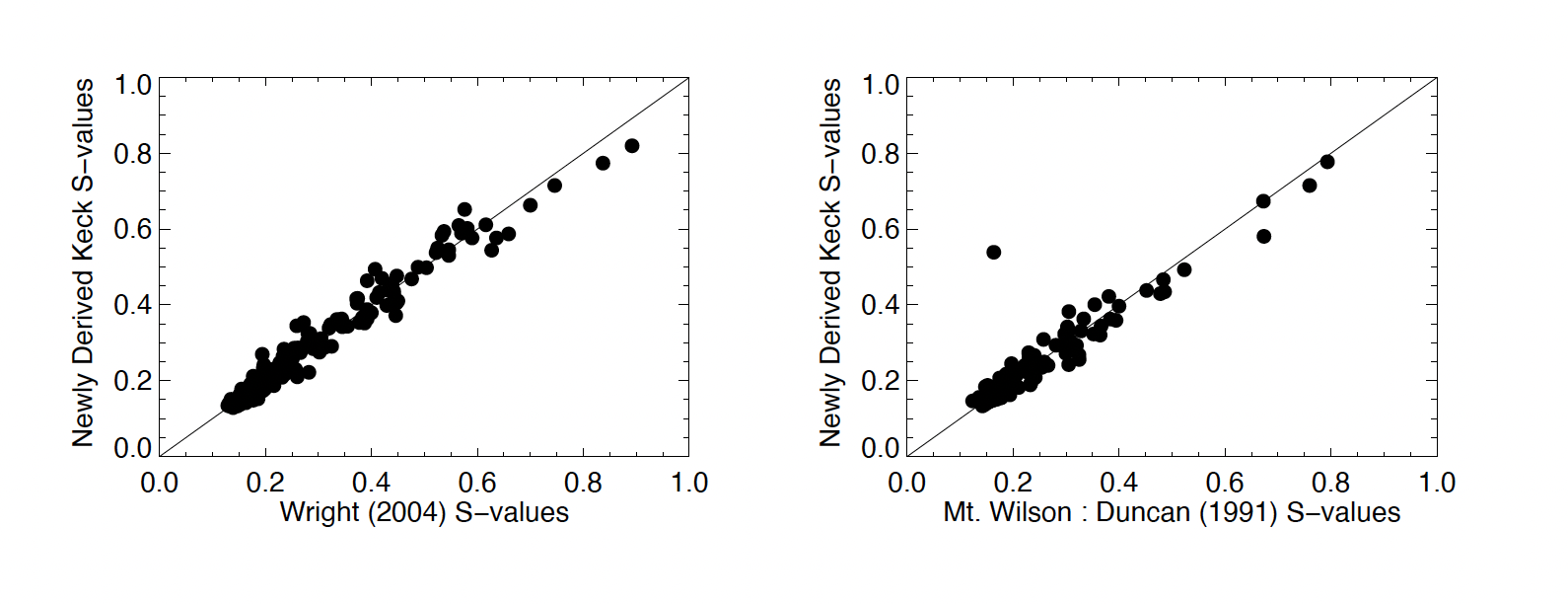}
\caption{By selecting stars that have been observed with HIRES by CPS and by the Mt. Wilson survey, we can determine the coefficients needed to convert flux values to S-values. 
 Left) We assess our S-values for consistency by comparing our newly created values with those published in \cite{Wright2004}, whose values were calculated from spectra collected on the previous HIRES detector, pre-2005. The S-values on the y-axis were created with the coefficients determined by fitting our newly extracted flux values to the Wright S-values. 447 stars were used in this comparison. The standard deviation of the residuals 0.020.  
 Right) The Mt. Wilson values from \cite{Duncan1991} are plotted against our newly determined values for 154 overlapping stars. Scatter in the relation is due to observing stars at different points in their activity cycles as well as imperfect accounting of the blaze function. The star at [0.2, 0.6] is a known outlier, HD 137778, detailed in \cite{Wright2004}. If we include HD 137778, the standard deviation of the residuals 0.044. Removing the single outlier reduces the standard deviation to 0.024. 
}
\label{fig:cal_mtwilson}
\end{figure*}

The newly created HIRES S-values are plotted against the \cite{Duncan1991} S-values, showing a standard deviation of the residuals of 0.023  (Figure \ref{fig:cal_mtwilson}, right panel). This is comparable to the \cite{Isaacson2010} S-values, that showed a scatter of 11\% when calibrated to the Mt. Wilson values. We attribute the larger scatter in the 2010 work to the broader range of stars in that sample, both in terms of activity and \teff, compared to this work. We verify our calibration by comparing our new S-values of 447 non-Kepler stars that also have S-values in the \cite{Wright2004} sample of planet search stars observed on HIRES prior from 1995 to 2004. The standard deviation of the residuals is 0.020 (Figure \ref{fig:cal_mtwilson}, left panel). 
We adopt 0.02 as the calibration uncertainty, similar to survey to survey calibration uncertainty found by \cite{Mittag2013}. 

\begin{figure*}
\includegraphics[width = 2.2\columnwidth]{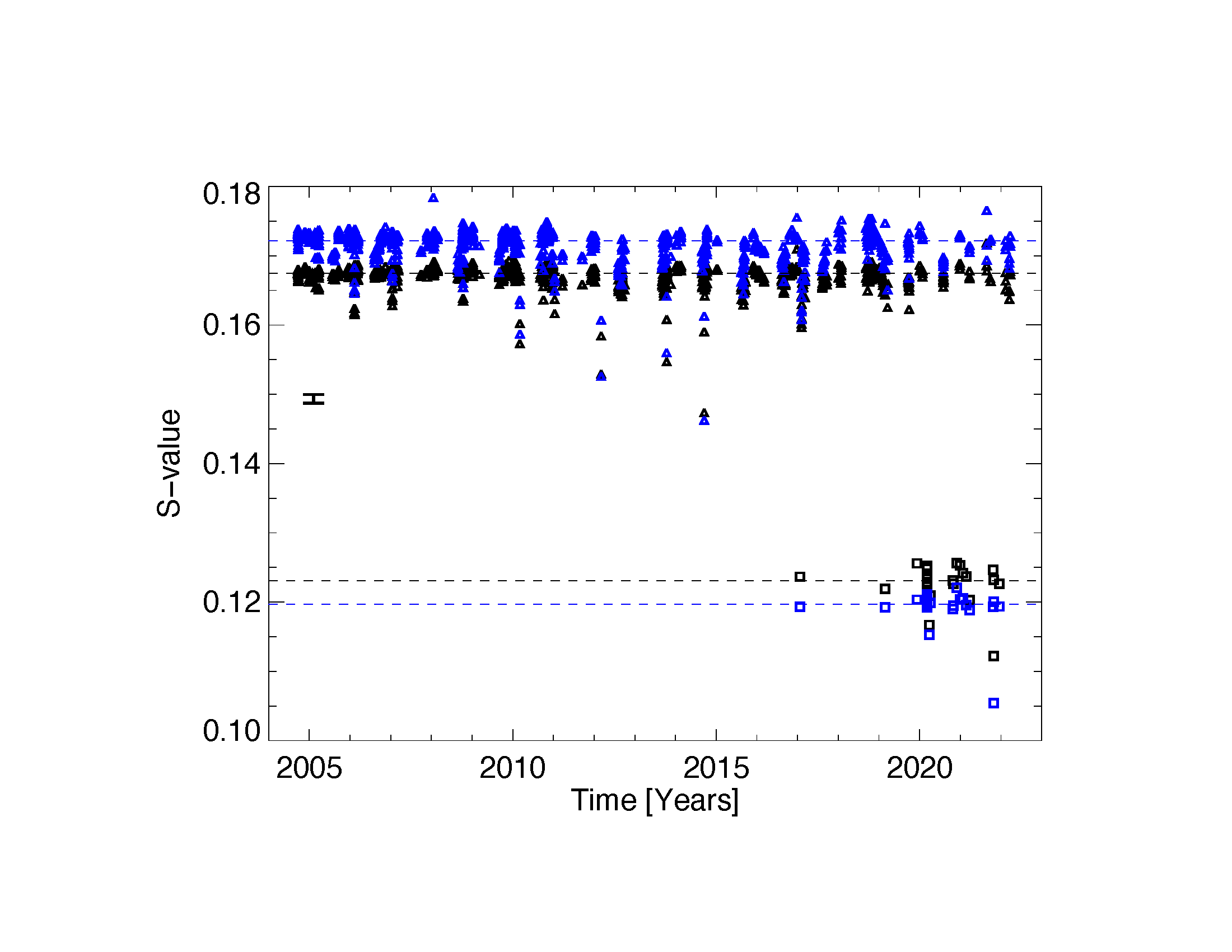}
\caption{ 
For the standard stars Tau Ceti (top set, triangles) and HD 60532 (bottom set, squares), S-values are calculated using the \cite{Isaacson2010} routine (black) and from the new HIRES-NSO routine (blue). The average values of 0.1670 (\rphk = -4.980) and 0.172 (\rphk = -4.954) respectively, are plotted as dashed lines. The offset of 0.005 or 3\% is due to different methods for addressing the blaze function as well as calibration errors. (The Mt. Wilson average value for Mt. Wilson is 0.175, compared to our HIRES-NSO median value of 0.172.) The standard deviations of the \cite{Isaacson2010} values and the newly derived values for Tau Ceti are 0.0016 and 0.0024. For HD60532, they are 0.0029 and 0.0031. The single epoch uncertainty of 0.03 is depicted at [2005,0.015]. We use 942 observations of Tau Ceti. A representative error bar 1\% as determined by the scatter of Tau Ceti and HD 60532, two S-value standards, is shown at [2005,0.15].}
\label{fig:svals_10700}
\end{figure*}

We verified continuity in the activity scale from the \cite{Isaacson2010} method to the new NSO method described in the paper by plotting both sets of S-values for two stellar activity standards, Tau Ceti and HD 60532 (Figure \ref{fig:svals_10700}). \cite{GomesdaSilva2021}, notes that the chromospheric standard star Tau Ceti varies by 0.83\% or a dispersion of 0.0015, and recommends using HD 60532 as well, with a scatter of only 0.36\%, and absolute dispersion of 0.0005. This tiny variation over time makes HD 60532 an excellent standard star for checking single instrument \cahk S-value precision.

There is a small offset between the values from the \cite{Isaacson2010} method and our new HIRES-NSO method that is visible for both standard stars. With the HIRES-NSO method, Tau Ceti shows a median S-value of 0.1720 with a standard deviation of 0.0023. The HIRES-NSO derived S-values for HD 60532 show a median value 0.1197 with a dispersion of 0.0031. For Tau Ceti, the \cite{Isaacson2010} method yields a median of 0.1670, with a standard deviation of 0.0016. For HD 60532 the median is 0.1231 with standard deviation of 0.0029. The new S-values for Tau Ceti and HD 60532 are offset by +0.005 and -0.004, respectively. 
The offsets are due to scatter in the calibration and the amplitude of the offset is similar to the amplitude of the scatter. Our uncertainties are consistent with previous works. For example, Tau Ceti shows very little long-term variation but calibration uncertainties result in a published values of 0.168 from \cite{Wright2004} and 0.175 from \cite{Duncan1991}. The performance of these standard stars gives us confidence in our extraction method and quantifies the error in our calibration.

Our distribution of S-values extends to lower values than previous surveys, with a tale toward very low activity that reaches \rphk $= -6.0$ 
(Figure \ref{fig:hist-doubleg}). We choose a value of \rphk $= -5.50$ as the lowest value that should be considered reliable. In Section \ref{sec:mm_search} we discuss very inactive stars and WASP-12 is an example of a star with the exceptionally low value of \rphk of $-5.50$. The low value is thought to be due to absorption beyond the chromosphere of the star \citep{Fossati2013}. Stars in our sample with \rphk values below $-5.50$ should be considered very-inactive, but these values likely underestimate their activity.

For the CKS-HK sample, most stars have a single observation, limiting our knowledge of the long-term behavior of these stars. For stars with two observations, we choose the one with higher SNR. For stars observed more than twice, such as those with multiple epochs for RV follow-up, we choose the median S-value. Time series S-values for these stars are available in \citep{Weiss2024}. The coolest stars in our sample have spectra that differ most significantly from the sun, and we find that the S-values for these stars are still meet our quality standards. Another limitation of using a single epoch in time to measure stellar activity is that we measure the activity at an unknown phase of the rotation period and activity cycle. This challenges our search for stars that are exceptionally inactive, in Maunder Minimum or magnetic minima states (Section \ref{sec:mm_search}. We rely on the statistical power, rather than time series, of our sample for the analysis of fundamental stellar properties, activity, and rotation periods. Without time series spectra to monitor long-term activity cycles, we are unable to make comparisons of years long cycles with rotation periods such as \cite{Brandenburg2017} and \cite{Metcalfe2016}. Instead, our analysis is similar to \cite{Zhang2020}, which also utilizes single epoch spectroscopy to assess activity for 59,816 stars. Our smaller sample has a \teff uncertainty of 60K vs 100K for the Large Sky Area Multi-Object Fibre Spectroscopic Telescope (LAMOST) sample \citep{Petigura2017}, and we focus our attention on the rotation periods from \cite{David2021}.

\subsection{The CKS-HK Sample}
\label{sec:sample}

The CKS sample has several components including a magnitude limited sample, with a cutoff at a \emph{Kepler} magnitude of 14.2, and a collection of fainter stellar host stars that includes habitable zone planets \citep{Borucki2013},  multi-planet systems \citep{Lissauer2014,Rowe2014}, and ultra-short period planets \citep{Sanchis-Ojeda2014}. The habitable zone planets and the multis fainter than 14.2 tend to have lower signal to noise ratios and are often omitted from our sample for having insufficient data quality in the bluer wavelengths where the \cahk  lines reside. To define the CKS-Gaia sample \cite{Fulton2018} make a similar magnitude restriction because the primary sample selection for stars fainter than V$\sim$14.2 was non-uniform. We choose to begin with the sample of 1189 planet stars hosting 1896 total planets from \cite{Fulton2018} and make additional quality cuts for our analysis. We did not include the stars from the recent CKS-Cool project \citep{Petigura2022} because the target SNR is too low at 4000 \AA\ to precisely measure the \cahk line fluxes.

In order to ensure sufficient quality of the S-value activity metric we made restrictions on the SNR near the \cahk  lines and on the local seeing conditions for each observation. The CKS-Gaia SNR for HIRES spectra was chosen to be $\sim$ 40 per pixel at 5000 \AA. This choice impacts the S-values resulting for cooler stars that have lower SNR in the \cahk  region compared to hotter stars due to inherent differences in their blackbody spectra. Beginning with the CKS-Gaia results (Table 1 from \cite{Fulton2018}) which has 1189 planet host stars, we remove 180 observations that have SNR less than five per pixel in the continuum regions near the \cahk lines. In most cases we use the same spectra as the CKS-Gaia project but for 25 stars higher SNR spectra, collected more recently, were available. From the 2D echellogram, we measured the seeing value for each observation and removed observations with seeing greater than 1.6\arcsec{}, ensuring high quality measurements \cite{Baum2022}. The HIRES spectrometer, a slit-fed spectrograph which uses an echelle grating as the cross dispersing optic, results in a 2D echelle format in which the orders become closer together toward bluer wavelengths. (spectrographs with cross-dispersing prisms have orders that are closer together in the red). Since we use the C2 decker (0.87\arcsec{} x 14.0\arcsec{}) for observations of faint stars in order to remove background sky flux \citep{Batalha2011}, when a faint star ($V >11$) is observed in poor seeing conditions, the bluest orders overlap causing cross order contamination and a poor quality S-value measurement. Removing 109 observations with poor seeing values and three stars that have no stellar mass or radius leaves 900 planet host stars with well characterized stellar properties and S-values. Three stars have no \logg value. Fourteen stars have an S-value lower than 0.10, chosen as a minimum value for calculating \rphk, leaving 879 host-stars. We define this as our CKS-HK stellar sample.

Binary star systems can challenge studies of stellar activity due to tidal interaction or spectral contamination, or with spectral contamination. To mitigate binary star contamination, stars with a detected secondary spectrum, with flux ratios as low as 1\% of the primary, using the technique of \cite{Kolbl2015} were identified by \cite{Fulton2015} as planet false positives and are excluded as such.  

The CKS-HK catalog of chromospheric activity is quite different compared to RV surveys of planet search stars.  While most RV surveys focus on either M-dwarfs or FGK stars providing large catalogs of high-resolution spectra, the stars in these surveys typically have an unknown number of short period planets with planet radii from 1-4 Earth-radii \citep{Rosenthal2021}. In comparison, every star in our sample has one or more known transiting planets, and the distance to the average \emph{Kepler} field star is about a kiloparsec rather than a 1-200 pc for typical RV survey stars. 

\subsubsection{The CKS-HK Planet Sample}

In addition to the quality metrics applied to the CKS-HK stellar sample, for analysis involving planet properties we make further qualifications. The CKS-HK planet sample will be defined by the quality cuts that are described in Section \ref{sec:sval_extract} and further quality cuts that depend on the planet properties in these systems. Much of the analysis focuses on planets smaller than 4.0 \rearth. We define this as our CKS-HK planet sample.

\subsection{Literature Data}

\subsubsection{\emph{Kepler} Stellar Rotation Curves}

The field of stellar rotation period analysis has richly benefited from the \emph{Kepler} 30-minute cadence with near continuous data collection for 90 days of a typical \emph{Kepler} quarter, up to four years over the life of the mission. While ground based photometric surveys had been critical in building our understanding of stellar rotation periods \citep{Henry1996,Duncan1991}, \emph{Kepler} has grown the number of available stellar rotation periods into the tens of thousands \citep{McQuillan2014}. 

Novel techniques applied to light curves can quickly analyze vast amounts of photometry. The auto-correlation functions (ACF) \citep{McQuillan2014} was used to create a catalog of rotation periods for 30,000 stars ranging from 0.2 days to 70 days across stellar masses from 0.1 to 1.3 \msun. \cite{Angus2018} used a machine learning technique, trained on that catalog, to determine rotation periods for \emph{Kepler} Objects of Interest (KOIs), which we use in Section \ref{sec:rotation_periods}. \cite{Santos2021} used a combination of wavelet analysis, ACFs and machine learning to further study the rotation periods of \emph{Kepler} stars. 

Recently, \cite{david2022} examined \emph{Kepler} rotation periods from \cite{Walkowicz2013,McQuillan2014,Mazeh2015,Angus2018} resulting in the identification of the "Rossby Ridge", a relationship between the \teff and stellar rotation period.  The Rossby Ridge results support the stellar spindown theory of weakened magnetic braking (WMB) as formulated in \cite{vanSaders2016}, which is a deviation from the spindown relationships that govern young stars until the ages of a few Gyr. We build upon vetted rotations periods of the CKS-Gaia sample from \cite{David2021}, adding the chromospheric activity measurements from the \cahk lines to explore \rphk and its relation to stellar rotation periods and stellar spin down in the Rossby Ridge in Section \ref{sec:rotation_periods}.

\subsubsection{The CKS-HK Stellar Rotation Sample}
\label{sec:cks-rotation-sample}

The CKS-HK rotation sample begins with the CKS-HK stellar sample, and is refined based on the quality of the determination of the photometric rotation periods from \emph{Kepler}. We use this sample to examine activity-rotation relations and stellar effective temperature, metallicity and stellar surface gravity. The stellar rotation analysis will require dividing the sample by stellar type, evolution and \feh. In order to utilize the most well-determined stellar rotation periods, we keep only the reliable stellar rotation periods from \emph{Kepler} photometry compiled and vetted by \cite{David2021} to define the CKS-Gaia sample of rotation periods. The quality of rotation periods are labeled 0, 1, 2, 3 as having no periodicity, an ambiguous period, a reliable period and a highly reliable period, respectively. With the CKS-HK stellar sample of 879 stars with valid chromospheric activity measurements with the vetted rotation periods to finalize our CKS-HK rotation period sample with 168, 325, 216 and 184 stars with reliability ranks of 0, 1, 2, and 3 respectively.

\section{Results and Discussion}
We begin our analysis by examining the relationships between our new \rphk  measurements of the chromospheric activity and the fundamental stellar properties from CKS-Gaia.

\subsection{\teff, \logg, \feh and \rphk}
\label{sec:act-stellar-props}

\begin{figure*}
\includegraphics[width = 2.0\columnwidth]{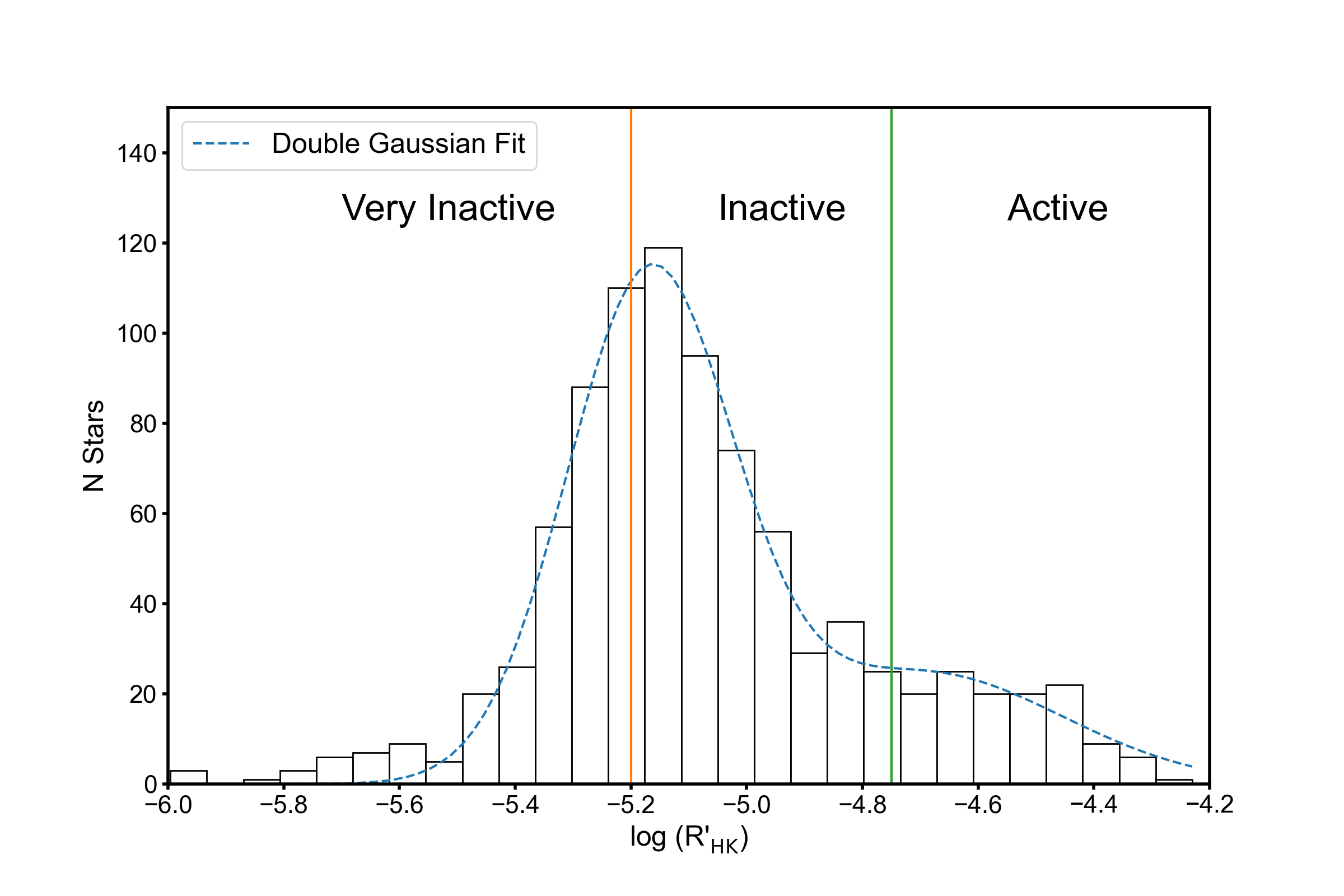}
\caption{ 
Double Gaussian Fit for the Full CKS Stellar sample of 893 stars.
} 
\label{fig:hist-doubleg} 
\end{figure*}

\begin{figure*}
\includegraphics[width = 2.2\columnwidth]{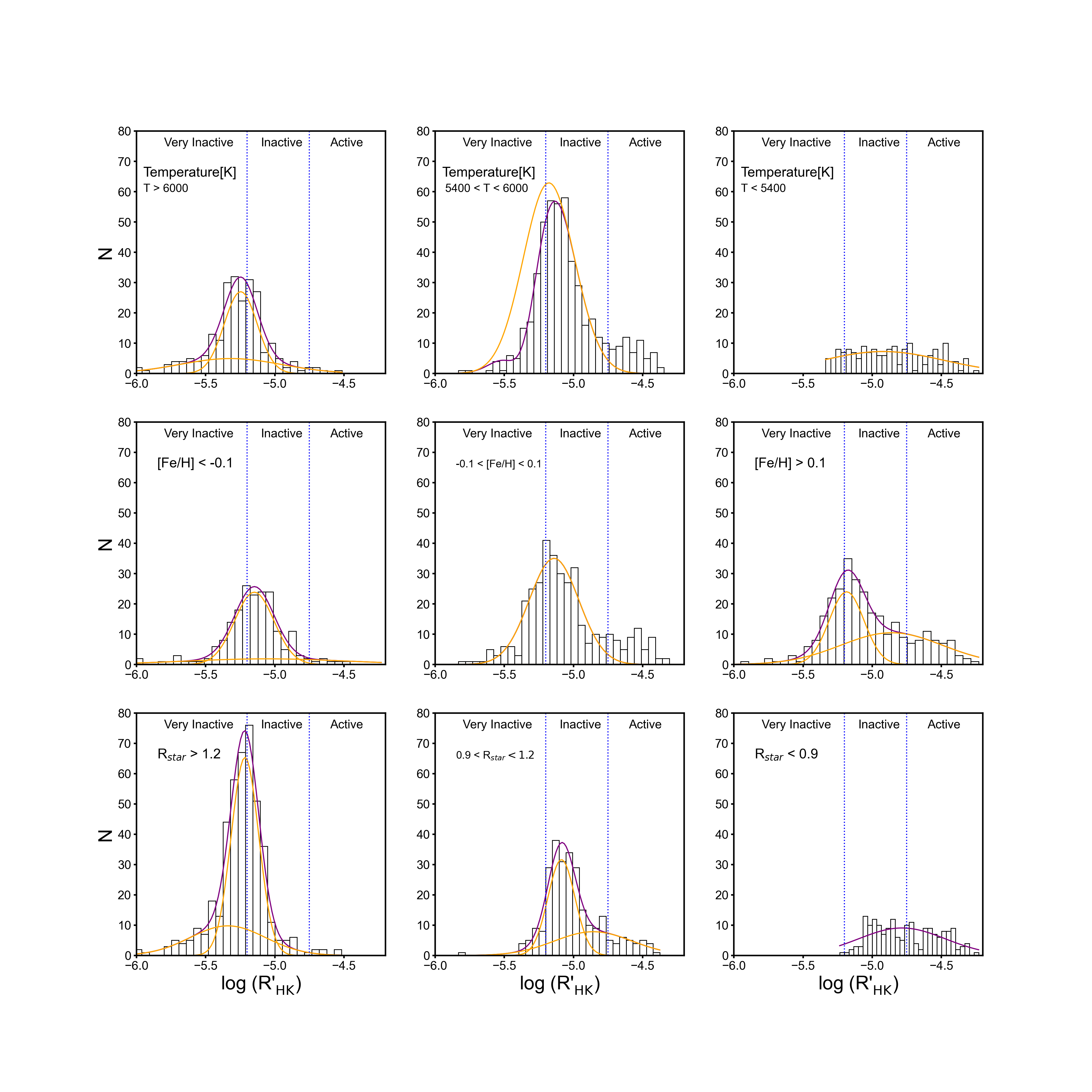}
\caption{ 
For each stellar property and property bin, the data are show in the histogram and the two model Gaussian fit is show in purple, with component Gaussians in yellow. A single component Gaussian fit was preferred for the coolest bin of \teff, the middle bin of \feh and the bin of smallest \rstar. 
Top row: \teff separates the three panels from left to right with break points at 6000 K and 5400 K. 
Middle row: \feh separated from highest to lowest with break points at -0.1 and +0.1. 
Bottom row: \rstar is plotted from the largest bin to the smallest with break points at 0.9 and 1.2 \rsun. The values of the fitted Gaussians are shown in Table \ref{tab:table_fits}. 
}
\label{fig:hist-9panel} 
\end{figure*}

We examine the full distribution of \rphk for our stellar sample, and model it with a two Gaussian fit (Figure \ref{fig:hist-doubleg}). as a function of the stellar properties \teff, \feh, and \rstar in Figure \ref{fig:hist-9panel}. To allow for quantitative comparisons between surveys (such as \cite{Santos2021}, we model three sub-sets of our sample as both single and double Gaussian distributions. The top panel of Figure \ref{fig:hist-9panel} shows the distribution of \rphk divided at \teff values of 6000 K and 5400 K, with the hotter F-stars on the left and cooler K-dwarfs on the right. In the middle panel we divide stars into bins of metallicity at +0.1 and -0.1 dex, with the most metal rich stars of the left and solar metallicity stars in the middle. The \rphk distribution as a function of stellar radii is divided at 0.9 and 1.1 \rsun with solar radius-like stars in the bottom middle panel. Gaussian fitting results for the full sample, and the sub-divided sample are compiled in Table \ref{tab:table_fits}.

\begin{table*}
\centering
\scriptsize
\caption{Gaussian Fit Parameters \label{tab:table_fits}}
\begin{tabular}{lrrrrrrrr}
\hline
      Property Bin & Amplitude & Mean & Sigma  & Amplitude & Mean & Sigma & Chi-squared & Reduced Chi-Squared \\

\hline
 Full Sample (893)    & 111.5 & -5.167 &  0.143 & 24.73 &-4.691 & 0.242 & 500.19 & 20.84 \\
 \teff > 6000         & 30.75 & -5.25 &  0.14 & ... &... &... & 309.30 & 11.46 \\
  5400 < \teff < 6000 & 56.89 & -5.12 &  0.14 &... &... &... & 825.83 & 30.59 \\
  \teff < 5400        &  7.26 & -4.92 &  0.44 &... &... &... & 180.21 &  6.67 \\
   \feh < -0.1        & 25.18 & -5.15 &  0.16 &... &... &... & 119.22 &  4.42 \\
   -0.1 < \feh < 0.1   & 35.03 & -5.14 &  0.18 &... &... &... & 796.18 & 29.49 \\
   \feh > 0.1         & 26.83 & -5.13 &  0.22 &... &... &... & 507.50 & 18.80 \\
 \rstar > 1.2         & 72.28 & -5.22 &  0.11 &... &... &... & 494.04 & 18.30 \\
 0.9 < \rstar < 1.2   & 35.02 & -5.07 & -0.13 &... &... &... & 436.10 & 16.15 \\
 \rstar < 0.9          &  9.08 & -4.78 &  0.31 &... &... &... & 267.61 &  9.91 \\

 \teff > 6000          & 27.00 & -5.25 &  0.12 &   4.94 & -5.32 & -0.37 & 248.36 & 10.35 \\
  5400 < \teff < 6000  & 62.91 & -5.18 & -0.18 & -28.45 & -5.34 &  0.11 & 697.72 & 29.07 \\
  \teff < 5400         &  7.26 & -4.92 &  0.44 &   1.00 & -5.70 &  0.03 & 180.21 &  7.51 \\
   \feh < -0.1         & 23.87 & -5.15 & -0.14 &   1.87 & -5.06 & -0.58 & 102.02 &  4.25 \\
   -0.1 < \feh < 0.1   & 14.38 & -7.06 &  0.05 &  35.03 & -5.14 & -0.18 & 796.18 & 33.17 \\
   \feh > 0.1          & 10.48 & -4.86 &  0.37 &  24.01 & -5.18 &  0.13 &  92.77 &  3.87 \\
 \rstar > 1.2          & 65.36 & -5.22 &  0.10 &   9.78 & -5.34 &  0.27 & 294.67 & 12.28 \\
 0.9 < \rstar < 1.2    &  7.84 & -4.86 &  0.28 &  31.62 & -5.09 &  0.10 & 198.27 &  8.26 \\
 \rstar < 0.9          & ...   &...    & ...   &...     &...    & ...   & ...    &  ... \\

\hline
\end{tabular}
\end{table*}

Sun-like stars are a common focus in activity analyses and our study focuses on stars between 4800 K and 6250 K. Stars with temperatures above the Kraft Break at \teff of 6250 K have thinning convective zones, dividing fully radiative and partially convective stars \citep{Kraft1967}. Fully radiative stars are magnetically different than those below the Kraft Break because they lack a tachocline that is thought to be the critical to the production of magnetic activity. The study of rotation and convection near the Kraft Break is an active field of study \citep{Metcalfe2019} and is relevant in our rotation period analysis. 
Generally, the stars with the largest radii in our sample are evolved and are also the least active. We will discuss the relationship between \feh and stellar rotation periods in Section \ref{sec:prot_feh}.

\begin{figure*}
\includegraphics[width = 2.0\columnwidth]{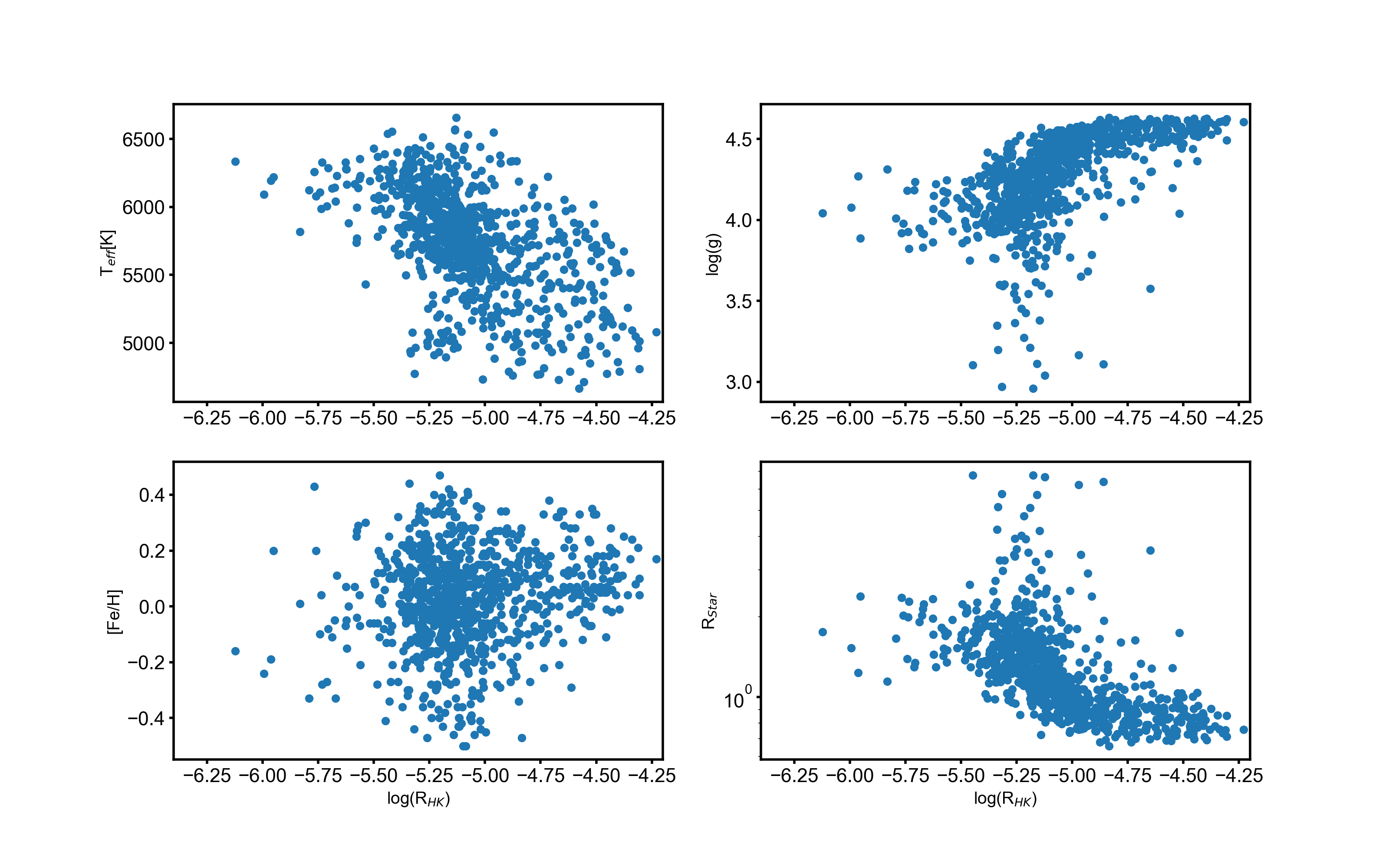}
\caption{ 
\rphk is plotted on the x-axis for all plots (more active stars are to the right) and fundamental stellar parameters are plotted on the y-axis. Top left: \rphk vs \teff shows that more active stars tend to be cooler. Top right: \rphk vs \logg reveals that sub-giants are sparse and are inactive (\rphk $<$ -5.1)  which are lower on the plot. Most main-sequence stars have a range of activities. Bottom left: \rphk vs \feh shows a balanced distribution with more active stars being more metal rich. This is consistent with those stars being younger. Bottom right: \rphk vs \rstar. This plot shows that as activity increases to the right, most stellar radii are near 1.0 \rsun } 
\label{fig:stellar_props_activity_4panel} 
\end{figure*}

\begin{figure*}
\includegraphics[width = 2.1\columnwidth]
{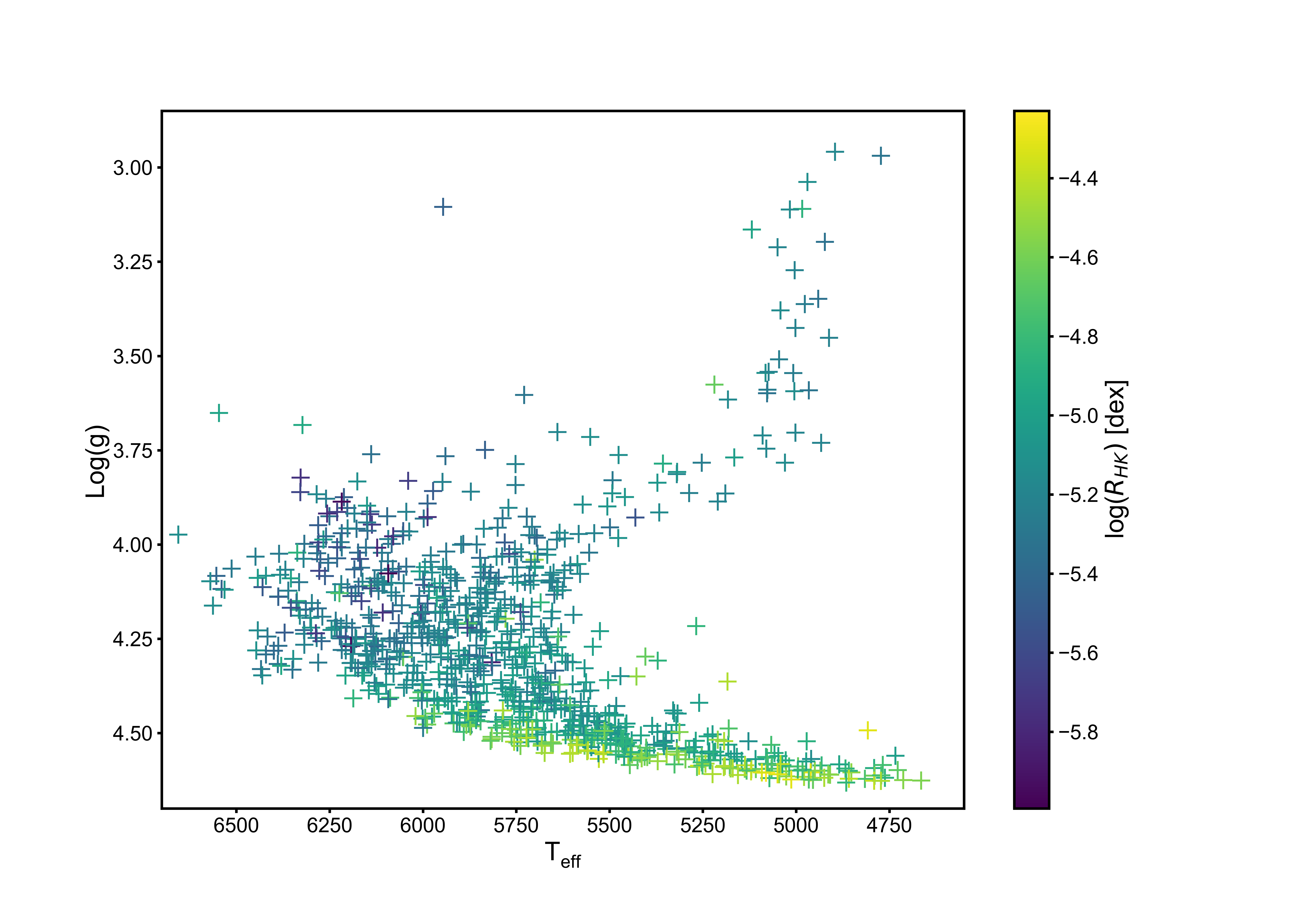}

\caption{Stellar surface gravity is plotted as a function of stellar effective surface temperature for the CKS-HK stellar sample of 879 stars. The color bar indicates \rphk  with yellow as more active and purple as less active. The giant branch moves from the center to the upper right of the plot. The most active stars (yellow) are cool dwarfs that make up the lower envelope of the main-sequence. From CKS-Gaia \citep{Fulton2018}, the typical error on \teff  and \logg  are 60 K and 0.01 dex respectively. We use the calibration error of 0.02 as the uncertainty on \rphk. }
\label{fig:teff_logg_rphk} 
\end{figure*}

We examine the CKS-HK stellar sample broadly in Figure \ref{fig:stellar_props_activity_4panel} by plotting the CKS-HK stellar properties sample as a function \rphk. In the temperature plot we see the most active stars are cooler than the sun, and all of the inactive stars have super-solar temperature. The relationships between stellar surface gravity and stellar radius relations to \rphk reveal the most active stars in our sample are on the main-sequence and near to 1.0 \rstar, consistent with Figure \ref{fig:teff_logg_rphk}. Viewing the sample in \teff vs. \logg space, the sub-giant population rises above the main-sequence as \logg decreases (Figure \ref{fig:teff_logg_rphk}). While the division between subgiant and main-sequence stars is ill-defined, we will use various cutoffs for \logg in the next few sections, including \logg = 4.0. The color scaling shows that most stars are in the Inactive or Very Inactive categories. The most-active stars exist along the lower envelope, with the highest \logg values, as expected for stars that are lie nearest to the zero-age main-sequence. For the rotation period analysis in Section \ref{sec:rotation_periods} and the study of the least-active (Section \ref{sec:mm_search}), we will focus on main-sequence stars rather than sub-giants.

When testing for a correlation between chromospheric activity and metallicity, we might expect to see that metal-rich stars, which tend to be younger than metal poor stars, and are also more active. This property is visible in the bottom left panel of Figure \ref{fig:stellar_props_activity_4panel} which shows a fairly smooth distribution around the average \rphk and \feh, with a slight overabundance of active stars that are metal rich. 

The least active stars are examined in detail in Section \ref{sec:mm_search} and we search for candidates for stars that are in Maunder Minimum or Magnetic Minimum (MM) type states \citep{Eddy1976,Saar2011}. Metallicity is also thought to be a factor in the study of stellar rotation periods as the metal content of the star can affect the depth of the convective zone, and therefore the convective turnover time and rotation period (See Section \ref{sec:prot_feh}). 

\subsection{Activity and Stellar Rotation Period}
\label{sec:rotation_periods}
Decades long observations of chromspheric activity measurements have identified relationships between chromospheric activity and stellar rotation, especially for solar-like stars. \cite{Noyes1984} formalized conclusions into equations that can be used to predict the rotation period based upon the average chromospheric activity of a star. By focusing on main-sequence sun-like, the Mt. Wilson studies were able to isolate variables such as stellar \logg, \teff (using B-V as a proxy), and to an unknown extent, \feh. \cite{Wright2004b} speculated that spectral synthesis and precise stellar abundances \citep{Valenti2005} would assist in finding very inactive stars.  Stellar activity in the least active stars revealed that changes in \logg due to stellar evolution plays is an important variable when identifying very inactive stars. \cite{Saar2011} showed that stellar metallicity has an impact on both the minimum activity of a star and on the stellar rotation period. Additional metal content in a star changes the opacity and is perhaps more important when the convective zone is thin, such as in stars near the Kraft Break. \cite{Amard2020A} used theoretical models with stellar masses of 0.8, 1.0 and 1.3 \msun and \feh between -0.5, and +0.5 and show that,  metallicity is indeed an important variable to consider when calculating rotation period. They confirm that the impact is more significant near the Kraft Break, where small changes in metallicity affect the convective turnover time and have a larger impact due to the thinner convective zone. 

In an observational test of those theoretical models \cite{Avallone2022} used \emph{Kepler} rotation periods, APOGEE data and Gaia parallaxes to analyze rotation periods for stars mostly hotter than 6250 K, the Kraft Break. They found no rotation dependence on metallicity but noted the difference in \teff for the two samples. The CKS-HK sample is well populated between 0.7 and 1.3 \msun and we analyze the metallicity in the range 0 +0.2 to -0.2 as well as, \teff,  \rstar and  \emph{Kepler} rotation periods in Section \ref{sec:prot_feh}. The metallicity dependence is explored along with other fundamental stellar parameters, and we consider the activity derived rotation periods in the same context. 

\subsubsection{Rotation Periods for the CKS-HK Sample}

Beginning with the CKS-HK rotation sample defined in Section \ref{sec:cks-rotation-sample}, we compare how the \cite{Noyes1984} rotation-activity relations used to calculate a stellar rotation period from \rphk  to  stellar rotation periods recovered from \emph{Kepler} photometry \citep{David2021}. We explore both methods of determining rotation periods and how they relate to the precise stellar properties from CKS-Gaia. Compared to the CKS techniques used to determine stellar properties to those used in \cite{Noyes1984}, we have much more powerful tools in form of high-resolution spectroscopy, to determine stellar surface gravity and metallicity. We also have a broader range of stellar temperatures which will expose the bias of solar-like stars when using the rotation periods derived from activity.

\begin{figure*}
\includegraphics[width = 2.1\columnwidth]{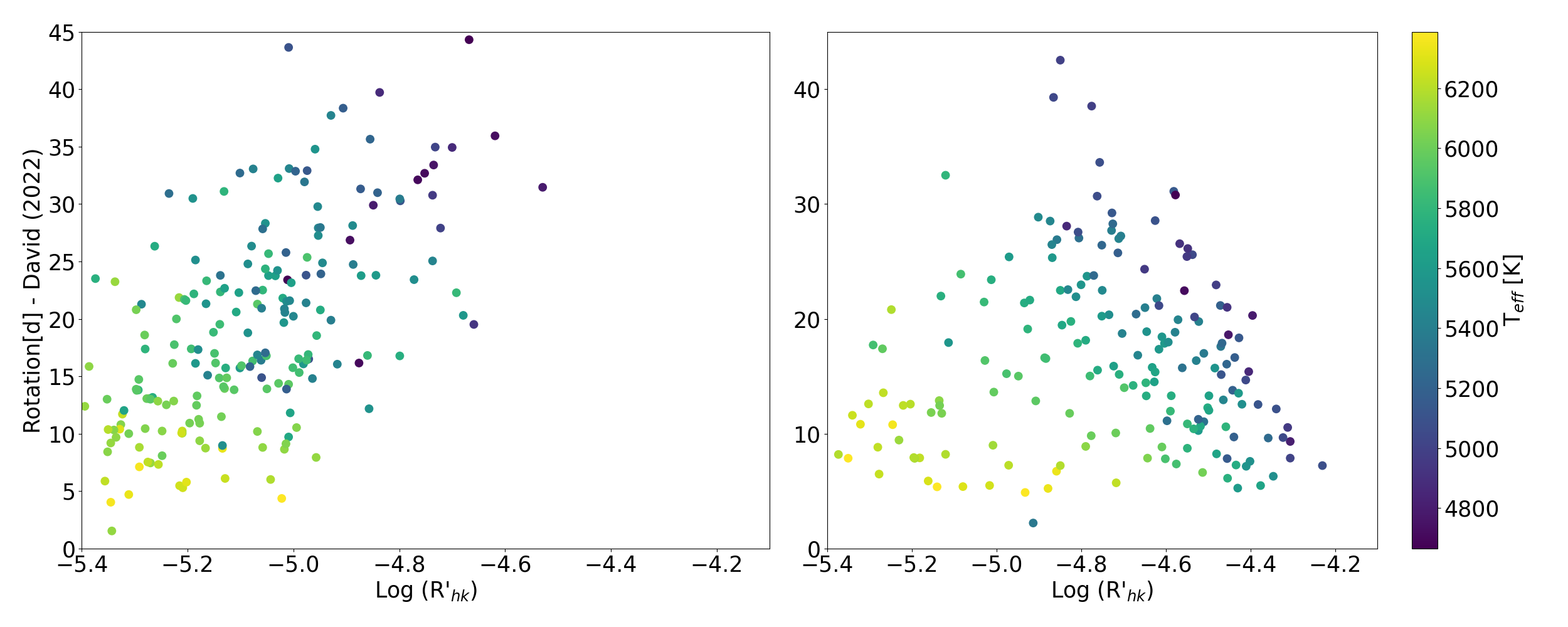}
\caption{Left panel: \emph{Kepler} rotation periods are plotted as a function of \rphk and all rotation periods from \cite{David2021} are included. Right: Only grade 3 rotation periods from \cite{David2021} are plotted.
In the left and right panel, 215 and 184 stars from the CKS-HK rotation sample are plotted, respectively. Recovering stellar rotation periods with single epoch \cahk  measurements is challenging because \rphk values for a given star naturally vary. Our sun's 11 year stellar activity cycle, which changes the average \rphk value  preserves the 27 day rotation period.   
The colors denote stellar surface temperature with cooler stars having darker shades. We expect rotation period to be related to activity within a given temperature bin. 
} 
\label{fig:rphk_Prot_david}
\end{figure*}

To visually confirm the relationship between activity and rotation, we plot \rphk versus the \emph{Kepler} rotation period in Figure \ref{fig:rphk_Prot_david}. As a function of \rphk we plot 215 stars with grade 2, 'reliable', stellar rotation periods from \cite{David2021} in the left panel and 185, grade 3, 'highly reliable' rotation periods in the right panel. The color scale shows stellar effective temperature from 4800 - 6400 K, the full range of the CKS-HK sample. If a strong correlation between rotation period and chromospheric activity is present, we expect stars with similar stellar temperature to have a smooth function with rotation period. Instead, in the left panel we see a large amount of scatter at every temperature where stars are less active.  By analyzing only stars that rank as \cite{David2021} highly reliable from the CKS-HK rotation sample, the temperature, rotation, activity relation is clarified (Figure \ref{fig:rphk_Prot_david}, right panel).  A correlation between stellar surface temperature, \rphk  and rotation period is now visible. \cite{Metcalfe2016} show a similar relationship (their Figure 1), noting the different slopes for different spectral types and lack of long rotation period stars for solar type stars.  We do not yet filter on stellar properties by removing sub-giants, but few subgiants have definitive rotation periods because they have fewer surface inhomogeneities. By focusing solely on the most well-determined rotation periods, and validated \rphk values, we can make further definitive statements about rotation periods and the stellar properties provided by the HIRES spectra in the CKS-HK rotation sample, which we do in Section \ref{sec:prot_stellar_props}.

\subsubsection{Photometric vs Activity Derived Rotation Periods}

Recovering stellar rotation periods with single epoch \cahk measurements is challenging because \rphk values for a given star naturally vary, similar to our sun's 11 year stellar activity cycle, changing the average \rphk  value while preserving the 27 day rotation period. \cite{Noyes1984} used a few tens of stars to create relations between Rossby number, convective turnover time and rotation period, but relied on average S-values of stars that had been collected over decades, which effectively averaged out the stellar activity cycles \rphk values. The long-term nature of that dataset makes it very valuable for analyzing stellar activity cycles as producing accurate long-term averages. We will utilize a larger number of stars with single epochs of activity measurements, and rely upon the larger number offset the lack of time series when we calculate rotation periods using the \cite{Mamajek2008} activity-period relations, which are very similar to \cite{Noyes1984}, and compare them with photometric rotation periods from \emph{Kepler}.

\subsection{Rotation Period, and Fundamental Stellar Properties}
\label{sec:prot_stellar_props}

\subsubsection{Overview}
To explain inconsistencies between stellar spindown models and observations of rotation periods for stars older than 1 Gyr, \cite{vanSaders2016}  developed theoretical relations that better describe spin down than empirical relations \citep{Skumanich1972}. \cite{vanSaders2016} showed that asteroseismically determined rotation periods \citep{Hall2021} align with their theory of WMB and  can reliably predict stellar rotation periods for stars older than one Gyr. 

The Van Sanders conclusions are reinforced by the work of \cite{Metcalfe2019}, who provide additional observational evidence that there exists a transition phase for stellar spindown that occurs in middle ages stars that leads to a breakdown of the previous spin-age relation for stars older than 1 Gyr. They show that the consistency between gyrochronology ages and ages determined with chromospheric activity break down near a value of \rphk of -4.95. In analyzing the rotation periods and \teff \cite{david2022} uses \emph{Kepler} derived rotation periods to solidify the conclusions of \cite{vanSaders2016} by showing that stars older than a few Gyr do not spin down beyond a certain point stalling to populate the Rossby Ridge.

\subsubsection{Rotation Period, \teff, \feh, and \rstar}
\label{sec:prot_feh}

\begin{figure*}
\includegraphics[width = 2.1\columnwidth]{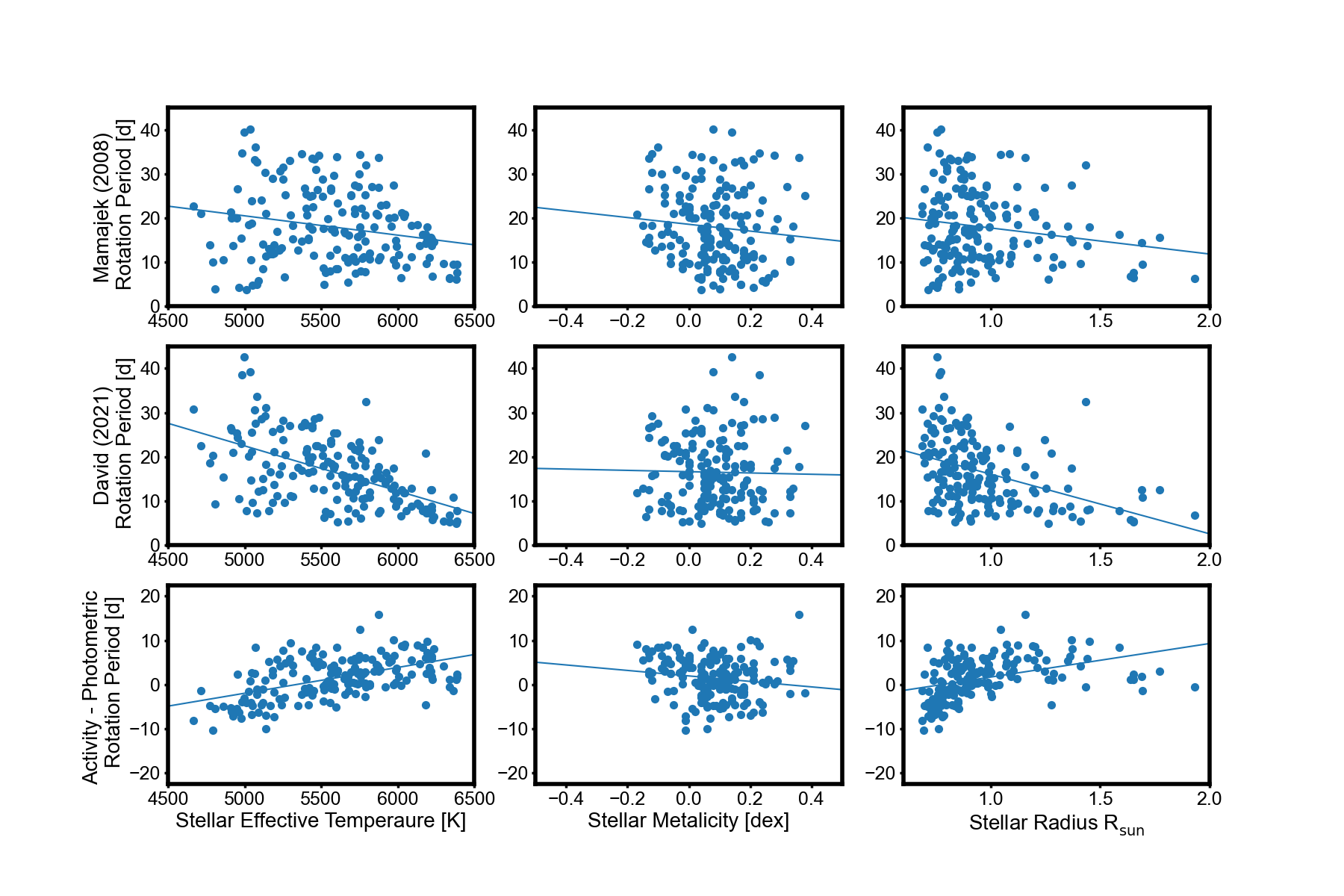}
\caption{ The top row shows the \cite{Mamajek2008} rotation periods, derived from S-values as a function of \teff, \feh\, and stellar radius from left to right. Each vertical column is the same in all rows. The second column shows the \cite{David2021} rotation periods with very-reliable grade. The bottom row shows the difference between activity derived rotation periods and the photometric rotation periods as a function of each stellar property for  of each plot for 173 stars from the CKS-HK rotation sample with \logg > 4.0 and grade 3 rotation periods. The rotation period uncertainties are 10\%. }
\label{fig:Prot_noyes_david_9}
\end{figure*}

We use the CKS-Gaia precise stellar parameters, our \rphk activity measurements, the activity-derived rotation periods, and photometric rotation periods to examine rotation-activity relationships. In Figure \ref{fig:Prot_noyes_david_9} we plot rotation periods derived from \rphk  using the \cite{Mamajek2008} equations, the \emph{Kepler} photometric rotation periods from \cite{David2021} and the difference in those values versus stellar properties for each star. We select only the grade 3, highly reliable, rotation periods from \cite{David2021} and we remove sub-giants by restricting the \logg to be greater than 4.1 leaving 173 stars. Using this highly selective set of rotation periods, we calculate the difference between \emph{Kepler} photometric rotation period and the activity derived rotation periods. 
For each panel in Figure \ref{fig:Prot_noyes_david_9}, we fit a linear trend to the data to assess trends in each stellar property. We checked for evidence that SNR of the S-value measurements affects the S-values themselves and found none, reducing the chance that a systematic error is occurring due to low SNR.

In the \teff\ vs $P_{rot}$ plots (Figure \ref{fig:Prot_noyes_david_9}, left column) we see the expected trend, hotter stars rotate faster, but the slope between the two rotation determinations is quite strong. If the difference plot for activity-derived rotation periods and photometric rotation periods showed a flat slope, then both determination methods would agree. Instead, the activity derived relation over-predicts rotation periods for cool stars and under-predicts rotation periods for hotter stars.  The main result of the stellar property and rotation period analysis that the relations for determining rotation period work very well for solar type stars, but less well as stars diverge from solar \teff. This reflects the bias toward solar-type stars used in the creation of past  activity-period relationships, and motivates the development of the WMB theory of stellar spindown.

When considering \feh (Figure \ref{fig:Prot_noyes_david_9}, center column), we see a linear correlation with metallicity for the activity derived periods, but no trend for the photometric periods. This disconnect may reveal the bias of Mt. Wilson survey toward solar metallicity stars.  The negative trend with stellar metallicity, shows that more metal rich stars have rotation periods that are under-predicted by activity. Contrast this with metal-poor stars showing rotation periods that are over-predicted compared to the photometric rotation periods. \cite{Amard2020A}, and \cite{Avallone2022} both studied the influence of metallicity on rotation periods. The former used theoretical models to conclude that stars of the same mass but with metal-poor versus solar metallicity rotate faster and have a higher Rossby number resulting from the change of the depth of the convective zone. The latter measured rotation periods in TESS and \emph{Kepler} light curves that do not show a statistically significant relationship between metallicity and rotation period. These analyses could be more definitive with larger samples of stars that have more extreme metallicities than we present. 

In Figure \ref{fig:Prot_noyes_david_9}, right column, the stellar radii are plotted against the activity-determined, photometric rotation period and the difference between the two. We see correlations in the two determinations of rotation periods which correlates with a faster rotation period for larger stars. The more numerous solar-type stars and sparse number of larger stars makes the conclusion ambiguous.

\begin{figure*}
\includegraphics[width = 2.1\columnwidth]{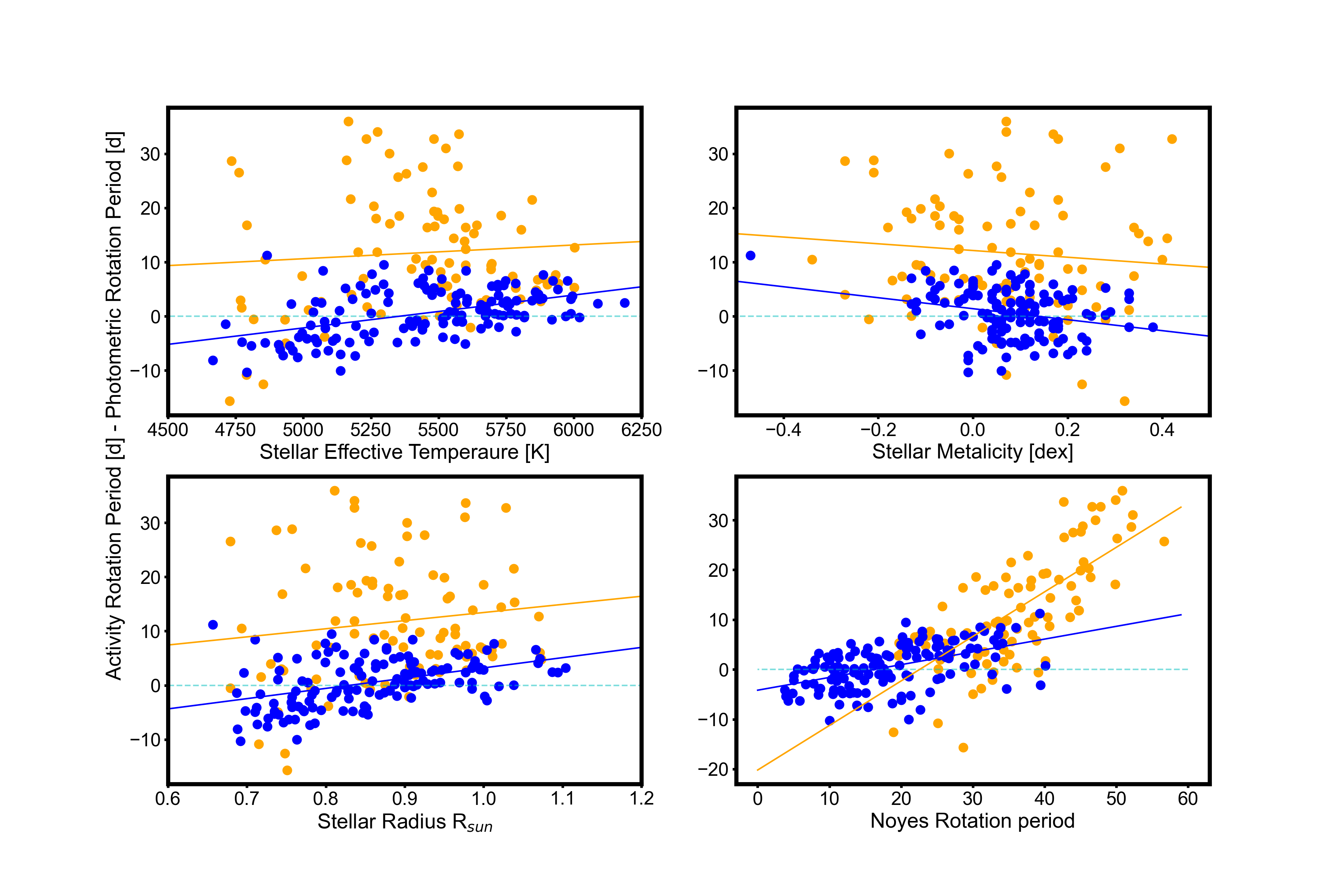}
\caption{The difference between photometric and activity derived rotation periods are plotted on the y-axis of each plot for 130 stars from the CKS-HK rotation sample with \logg > 4.4 and have rotation periods with grade 2, reliable (orange points) or grade 3, high reliable (blue points). The rotation period uncertainties are 10\%. Top left: The grade 2 rotation periods are far more discrepant than the grade 3 rotation periods and both show the same modest correlation with \teff. 
Top right: The grade 2 rotation periods show a larger variation in metallicity than those with grade 3. Bottom right: The grade 2 rotation periods show a strong correlation when plotted against the activity derived periods. This shows that for stars with a clear but not strong photometric rotation period (grade 2) then the discrepancy between the two methods of obtaining the rotation period is more than 30\% for periods beyond 40 days. If the photometric rotation period is very clear the agreement is much better, but the photometric rotation periods do not extend beyond 40 days. Bottom left: Stellar radius shows a modest slope, meaning there is a systematic error in rotation period determinations from activity metrics that is linear with \rstar. }
\label{fig:Prot_noyes_david}
\end{figure*}

In Figure \ref{fig:Prot_noyes_david} we plot only the difference panels from the previous figure and we now include grade 2 (reliable, orange points) and grade 3 (highly reliable, blue points) rotation period labels.  The differences between the activity-derived and photometric rotation periods are more pronounced for those that are deemed less reliably determined. This is likely explained by the lower precision ground-based data, compared to Kepler data, that was used to generate the rotation-activity relations from \cite{Noyes1984} and \citep{Mamajek2008}.

\begin{figure*}
\includegraphics[width = 2.1\columnwidth]{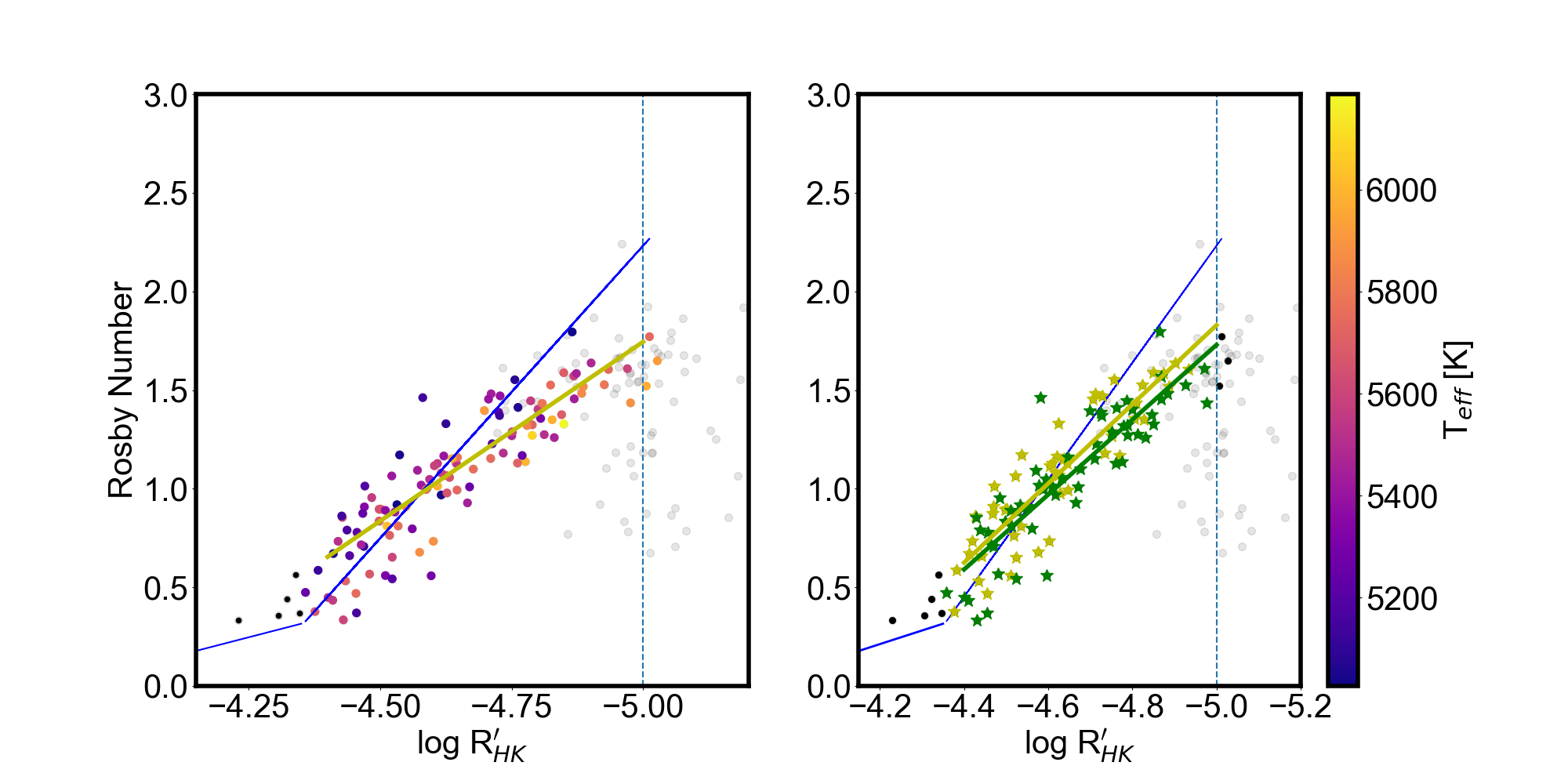}
\caption{Left panel: The CKS-HK rotation sample is shown in relation to their \rphk values and Rossby numbers as determined by the \emph{Kepler} rotation period divided by the convective turnover time as defined in \cite{Noyes1984}, equation 4. The best-fit from \cite{Mamajek2008}, equation 5 is overplotted in blue. \teff is less than 6250 K and \logg is required to be greater than 4.4 matching those cuts in that paper. The vertical line is the value of -5.0, shown to be the limit of the rotation activity relations. The colors scale is stellar surface temperature.
Right Panel: The two colors represent [Fe/H] values divided at the median of the distribution, \feh = +0.09. The yellow and green fits to the lower and higher metallicity data are consistent with each other, showing that the impact of metallicity on this relation is weak in the range of \feh: [-0.3,+0.3]. The grayed-out data points are the CKS-HK stellar sample that have grade 0 or 1 stellar rotation periods and are not used in the fits.
} 
\label{fig:rphk_rossby_mamajek7}
\end{figure*} 

\subsubsection{Rossby Number and Chromosperic Activity}

Recent work by \cite{david2022} used CKS-Gaia stellar properties and \emph{Kepler} photometric rotation periods to reveal the Rossby Ridge, a clustering of stars in \teff vs rotation period space that supports the theory of WMB. We use the CKS-HK rotation sample and our \rphk measurements to determine if chromospheric activity supports the presence of a Rossby Ridge. First we examine our \rphk values by comparing to the population of main-sequence stars used by \cite{Mamajek2008}.

The theoretical connection between chromospheric activity and stellar rotation is via the Rossby number, the ratio of the rotation period to the convective turnover time. By calculating the Rossby number using the \emph{Kepler} rotation periods, we can compare \rphk and Rossby number  (\citealt{Mamajek2008} their Figure 7). We focus our analysis in the active regime (-5.0 $\leq$ \rphk  $\leq$ -4.3 ), since we lack a sufficient number of stars in the very-active regime (\rphk $\leq$ -4.3).  In Figure \ref{fig:rphk_rossby_mamajek7} we plot \rphk vs the Rossby number for our CKS-Gaia stellar sample, using 107 'very-reliable', photometric rotation periods (grade 3). The temperature range has been limited from 5000K to 6200K and the \logg must be greater than 4.4, focusing on main-sequence stars and matching \cite{Mamajek2008}. Only data in the bounds of \rphk between -4.3 and -5.0 are used in the fit. The gray data are CKS-HK stars with unreliable, indistinct or 'reliable' rotation periods, flags of 0, 1, or 2 from \cite{David2021}. 

To assess subtleties in Figure \ref{fig:rphk_rossby_mamajek7}, we split the sample in metallicity to reveal that \feh does not play a strong role in the relationship between Rossby number and \rphk. We show the same data in both panels of Figure \ref{fig:rphk_rossby_mamajek7}, with the best-fit from \cite{Mamajek2008} in blue. The color bar describes the \teff of the sample in the left panel. A vertical dotted line at \rphk = -5.0 identifies value where we expect a breakdown in the standard activity-rotation relations \citep{Metcalfe2019}. By fitting a line to the data in the left panel we find a fit with a shallower, but similar slope compared to \cite{Mamajek2008}.  The right panel shows a similar version with the stellar metallicity highlighted as two distinct colors divided at the median value of those plotted, +0.09. We find a very similar slope when fitting the lower metallicity and higher metallicity halves of the data separately, suggesting that metallicity does not play a strong role in this relationship, at least in the range of [-0.2 to +0.3].

Figure \ref{fig:rphk_rossby_mamajek7} reveals an upper limit of 1.75 on the Rossby number. Theoretical work suggests that the critical Rossby number is 2.0 or 2.16, higher than any of our target stars. This is explained by our restrictions on rotation period quality and the gravity cutoff of 4.4. 

\cite{david2022} provides evidence in support of the  \cite{vanSaders2016} theory of WMB via the identification of the long period pileup in the temperature-rotation plane that results from magnetic braking. By using the CKS-Gaia sample and controlling for the quality of the rotation periods the Rossby Ridge is  identifiable.  We rely on the same robust stellar properties from the CKS-Gaia sample and strictly vetted rotation periods to further confirm the theory of WMB. 

\begin{figure*}
\includegraphics[width = 2.0\columnwidth]{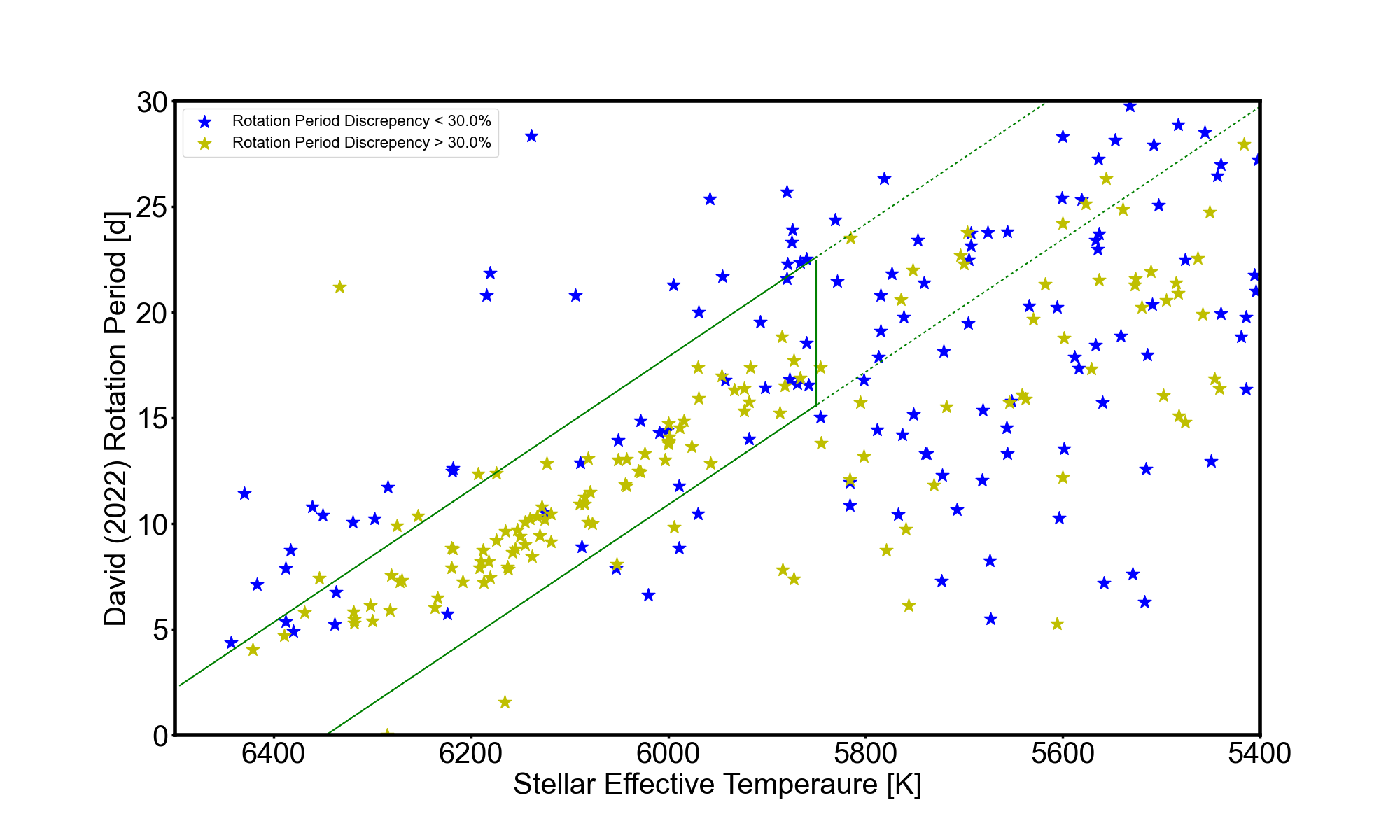}
\caption{The temperature v rotation period plane similar to \cite{david2022} (their Figure 4). Each symbol represents a star from the CKS-HK rotation period sample with grade 2 or 3 rotation period and \logg  > 4.0. Blue symbols identify stars that show agreement between their photometric and activity derived rotation periods to better than 30\%. Yellow data points show discrepancy larger than 30\%. Large discrepancy points to a disassociation of the stellar spin-down with age, consistent with the theory of weakened magnetic braking.
We extend the trapezoid defined in \cite{david2022} to 5600 K but find no evidence in the \rphk  data that inform the breakdown of the Ridge into that range of \teff.
} 
\label{fig:fig_teff_act3}
\end{figure*}

We replicate the temperature vs rotation period plot from \cite{david2022} and add chromospheric activity data in the form of rotation periods derived from \rphk values (Figure \ref{fig:fig_teff_act3}). We color code the stars for which the activity derived rotation period is larger than the photometric rotation period by more than 30\%.  With the standard uncertainty of the photometric rotation periods of 10\% according to \cite{david2022}, the choice of  30\% corresponds to a 3-sigma discrepancy between rotation period determination methods. For all stars in the \teff range of 5850-6250 K, the photometric rotation period is smaller than the activity derived period. We find that 79\% of the stars in the trapezoid region have over-predicted rotation periods of more than 30\%. This discrepancy points to the disconnect between the rotation period and magnetic field of a star as stars reach a few Gyr old, as described by the theory of WMB. \cite{Masuda2022} points out that the pile-up of stars in the P$_{rot}$ vs \teff plane is also affected by the decrease in the photometric amplitude of the rotation signal and that the pile-up may in fact be at longer periods but is currently below our detection limits.

\begin{figure*}
\includegraphics[width = 2.0\columnwidth]{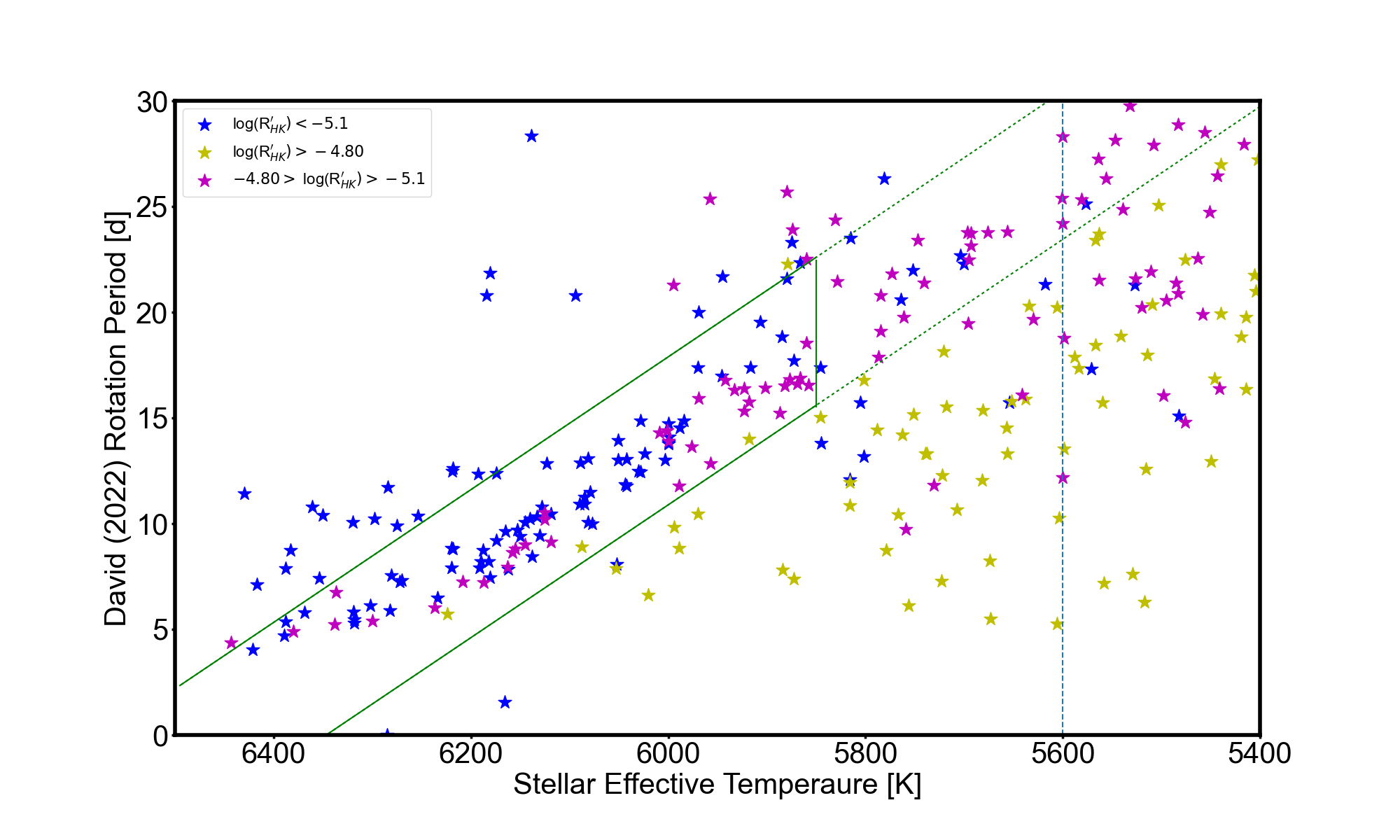}
\caption{The \teff vs rotation period plane similar to \cite{david2022} (their Figure 4). Each symbol represents a star from the CKS-HK rotation period sample with grade 2 or 3 rotation period and \logg > 4.4. Blue symbols are least active \rphk < -5.1, representing inactive stars. Yellow symbols have \rphk > -4.80, representing the active stars. Yellow symbols are within 0.15 of the critical value of \rphk = -4.95, where we expect the traditional spin down relations to break down in favor of the relations best described by weakened magnetic braking. 
} 
\label{fig:fig_teff_act2}
\end{figure*}

To further explore the \teff and rotation period plane, we change the color-code of \rphk values in Figure \ref{fig:fig_teff_act2} and search for a connection between the \rphk values and the Rossby Ridge. By dividing the \rphk values at -4.8 and -5.1, we can identify stars that lie along the Rossby Ridge. There are 3/103 stars with \rphk > -4.80, 34/103 between \rphk of -4.80 and -5.10 and 66 stars in the least active category of \rphk < -5.10. We expect stars of a certain \teff, say 6000 K to slowly spindown with age but to not change temperature while on the main- sequence. If we had a population of more quickly rotating stars that populated the area below the trapezoid, we could likely see the most active stars with the fastest rotation periods, and stars with ever decreasing activity as they approach the Rossby Ridge. Our population of stars from 5850 K to 6200 K has insufficient younger, and more active stars to see that type of behavior which is seen in surveys with larger numbers of stars such as \citep{McQuillan2014}.

We extend the trapezoid defined in \cite{david2022} from 5850 K to 5600 K and find tentative evidence in the \rphk data that indicates Rossby Ridge projects into cooler ranges of \teff. These stars have thicker convective zones, spend longer on the main-sequence, and take longer to spin down. These larger parameter spaces, and longer time spans in stellar evolution are a barrier to theoretical studies attempting to explain spindown for broader stellar populations. The conclusions by \cite{Masuda2022} regarding the amplitudes of stellar rotation signals are even more relevant for this range of \teff since these stars are inherently fainter, further clouding our ability to draw conclusions for these stars. 

\subsection{Ages and Chromospheric Activity}
\label{sec:ages}

The works of \cite{david2022}, \cite{vanSaders2016} and \cite{Metcalfe2019} attempt to find a relationship between age, activity and rotation, similar to  \cite{Noyes1984}, but with modern tools. \cite{Noyes1984} used rotation periods and a few assumptions about stellar structure.  By examining two stars of similar \teff\ and near the Kraft Break \cite{Metcalfe2019} reveal the age activity-relation discontinuity, near \rphk\ of -4.95. For more active stars the relationship between chromospheric ages and gyrochronology ages agree, but once stars' activity decreases beyond \rphk\ of -4.95, the age determinations diverge.  

\begin{figure*}
\includegraphics[width = 2.1\columnwidth]{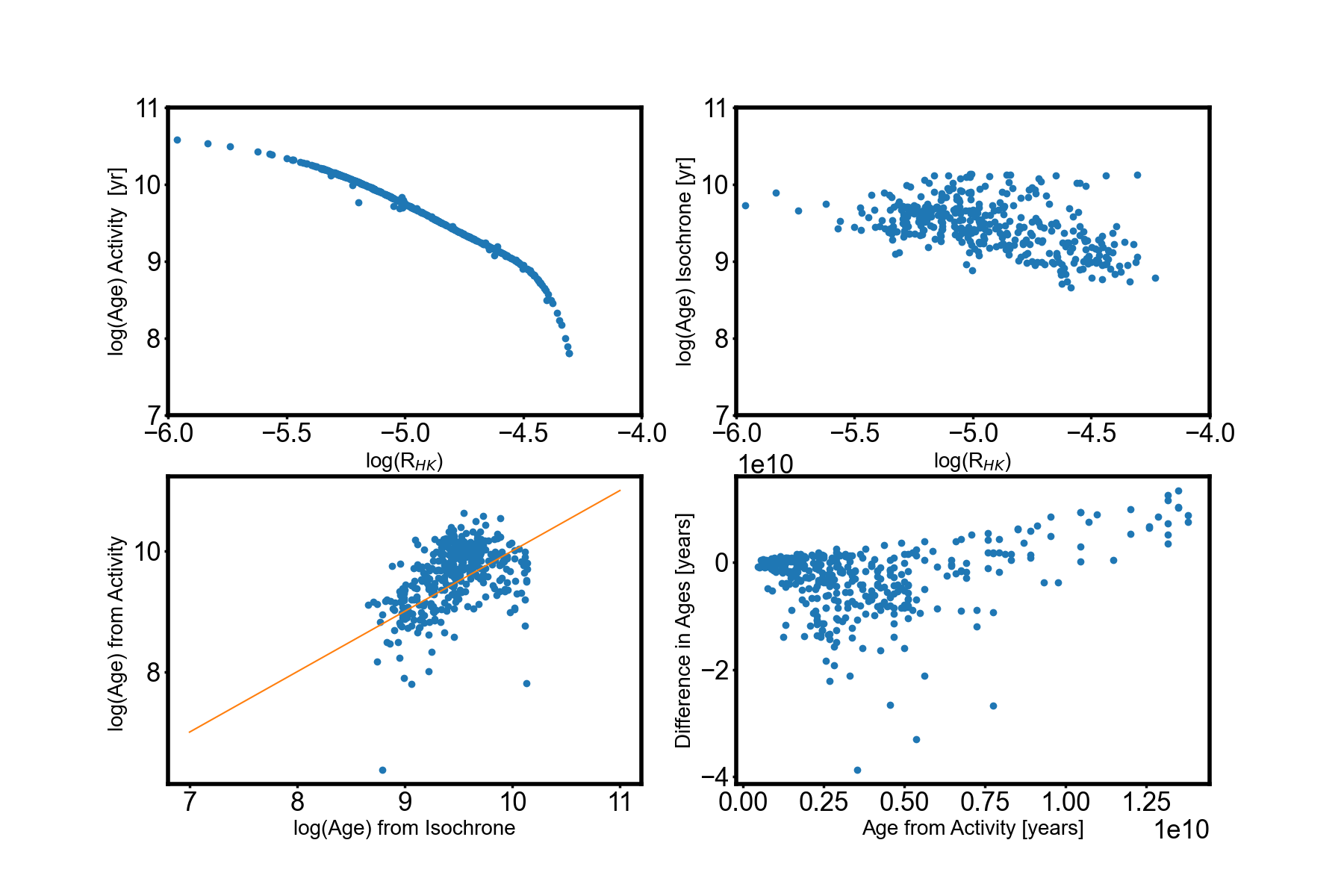}
\caption{We explore the age vs activity relationship with these four panels. Top left) The predicted rotation period determined the \rphk value of the star. Top right) The isochronal age from \citep{David2021} or CKS-Gaia as a function of the chromospheric activity. Bottom left) The isochronal age compared to the activity derived age and the one-to-one line overplotted. Bottom right) The difference between the two age determinations as a function of the isochronal age. Note this plot is in units of years, and the others are in log (age).
} 
\label{fig:fig_rphk_age_4panel}
\end{figure*}

Figure \ref{fig:fig_rphk_age_4panel} shows our two age metrics as function of \rphk in the two top panels. In the two bottom panels, the isochronal age is compared to the activity-derived age, and to the difference in the two age metrics for each star, in the left and right panels respectively.

Broadly speaking, the chromospheric ages are over-predicted compared to the isochrone ages. Given our evidence in favor of WMB, we expect the activity-age determinations to perform poorly for stars older than 1 Gyr. However, the mismatch in ages between the youngest, most active, stars is unexpected. The S-values and \rphk values for the CKS-HK star sample are collected in Table \ref{tab:table1}.

\begin{table*}
\scriptsize
\caption{Activity metrics and derived values. The Full Table has 1189 Entries}
\label{tab:table1}
\begin{tabular}{llrrrrrrlrlr}
\hline
Starname &S-value&\rphk &SNR         &$P_{Rot}$  &     log(Age) &Quality&Quality    &$P_{Rot}$ &$P_{Rot}$ \\
(KOI) &         &       &per pixel   &Activity[d]& Activity [yr]&       & Flag  &Phot[d] & Flag  \\

\hline
     K00001 &  0.145 &  -5.101 &   8 &      30.1 &   9.90 &       0 &   bad seeing &      0.0 &             1 \\
     K00002 &  0.128 &  -5.247 &  11 &       6.9 &  10.09 &       1 &           ok &      0.0 &             1 \\
     K00006 &  0.140 &  -5.103 &  19 &       8.5 &   9.91 &       1 &           ok &      0.0 &             0 \\
     K00007 &  0.141 &  -5.136 &  27 &      30.9 &   9.95 &       1 &           ok &      0.0 &             1 \\
     K00008 &  0.202 &  -4.779 &  14 &      16.6 &   9.40 &       1 &           ok &      0.0 &             0 \\
     K00010 &  0.127 &  -5.270 &  19 &      16.8 &  10.12 &       1 &           ok &      7.5 &             2 \\
     K00017 &  0.134 &  -5.202 &  10 &      41.8 &  10.04 &       1 &           ok &      0.0 &             1 \\
     K00018 &  0.125 &  -5.289 &  19 &      10.9 &  10.14 &       1 &           ok &      0.0 &             1 \\
     K00020 &  0.136 &  -5.178 &  21 &      27.3 &  10.00 &       1 &           ok &      0.0 &             1 \\
     K00022 &  0.135 &  -5.190 &  17 &      29.5 &  10.02 &       1 &           ok &      0.0 &             1 \\
\hline
\end{tabular}
\end{table*}

\subsection{The Most Inactive Stars: Overview} 
\label{sec:mm_search}

Stars that mimic the activity of the sun during its persistent minimum in the 17th century have been sought in an attempt to solidify the solar-stellar connection. \cite{Duncan1991} collected 18 years of S-value time series with the Mt. Wilson H\&K Activity project providing many examples of stars with solar-like stellar activity cycles \cite{Baliunas1995}, and several candidates for stars in low activity states. These stars were chosen as bright, sun-like stars, but without the vigor of spectroscopic stellar classification that we have today. 
One method for identifying a star in Maunder Minimum like state (or Magnetic Minimum(MM)) is to look for a decrease in the S-values over time or a transition from a stellar cycle into a non-cycling state.  Several candidate Maunder Minimum stars from the Mt. Wilson survey data alone were identified, though none has stood up to further scrutiny \citep{Wright2004b} without additional data.
Continued efforts to add modern data to the Mt. Wilson HK project's timeseries data have identified several MM candidates. 

With the addition of California Planet Search \rphk data, \cite{Shah2018} identified HD 4915 as a candidate for being in an activity minimum, but recent data showed the activity minimum is ending. Another analysis of the Mt. Wilson HK project plus CPS data \citep{Wright2004,Isaacson2010} extended the observing baseline for 59 stars from two to five decades. 

\cite{Baum2022} curated the various data sets and identified HD 166620 as a star that was in a previously cycling state and is now in a non-cycling low-activity state.  An extended effort has confirmed HD 166620 as a true MM star \cite{Luhn2022} using photometry and critically timed S-value measurements. In that study, intraseasonal variability was considered a more useful diagnostic of a MM state than instantaneous measurements that are anomalously low.  

Searching for MM stars by analyzing decades of time series data is laborious and time consuming so other methods for identifying stars in activity minima have been explored. Surveys like the Hipparcos Mission \citep{Perryman1997} made the identification of outliers more robust with precise measurements of parallax. The distance can be combined with spectral surveys that provide precise stellar surface gravities and metallicities \citep{Wright2004b,Fulton2018,Rosenthal2021}. Naturally, Gaia now provides precise parallax measurements for all stars down to V$\sim$19. The parallaxes can help to disentangle inactive main-sequence stars from evolved stars that are naturally inactive. The Gaia spectral measurements of the Calcium infrared triplet may enable statistical searches for inactive stars based on a small number of measurements for each star.

We can use the CKS-HK sample of stars to identify the least active and present candidates of stars in magnetic minima. 

\subsubsection{Identifying Stars in States of Magnetic Minima via Stellar Properties}

The discussion on what constitutes a MM candidate has become more nuanced as our ability more precisely measure the fundamental stellar parameters of surface gravity, effective temperature and metallicity have improved. We attempt to identify stars in a MM state, using a variety of methods including those of \cite{Wright2004} and \cite{Saar2011}. \cite{Henry1996} claim that using a single \rphk value to identify quiet stars, for example with an \rphk < -5.1, is insufficient. Using \teff, \logg  and \feh\ is required \citep{Saar2011}, and multi-epoch spectroscopy 
 is ideal \citep{Baum2022}. The CKS-Gaia stellar properties catalog enables a detailed study of activity, temperature, surface gravity and metallicity in the search for the MM stars using single epoch spectra.

\begin{figure*}
\includegraphics[width = 2.1\columnwidth]{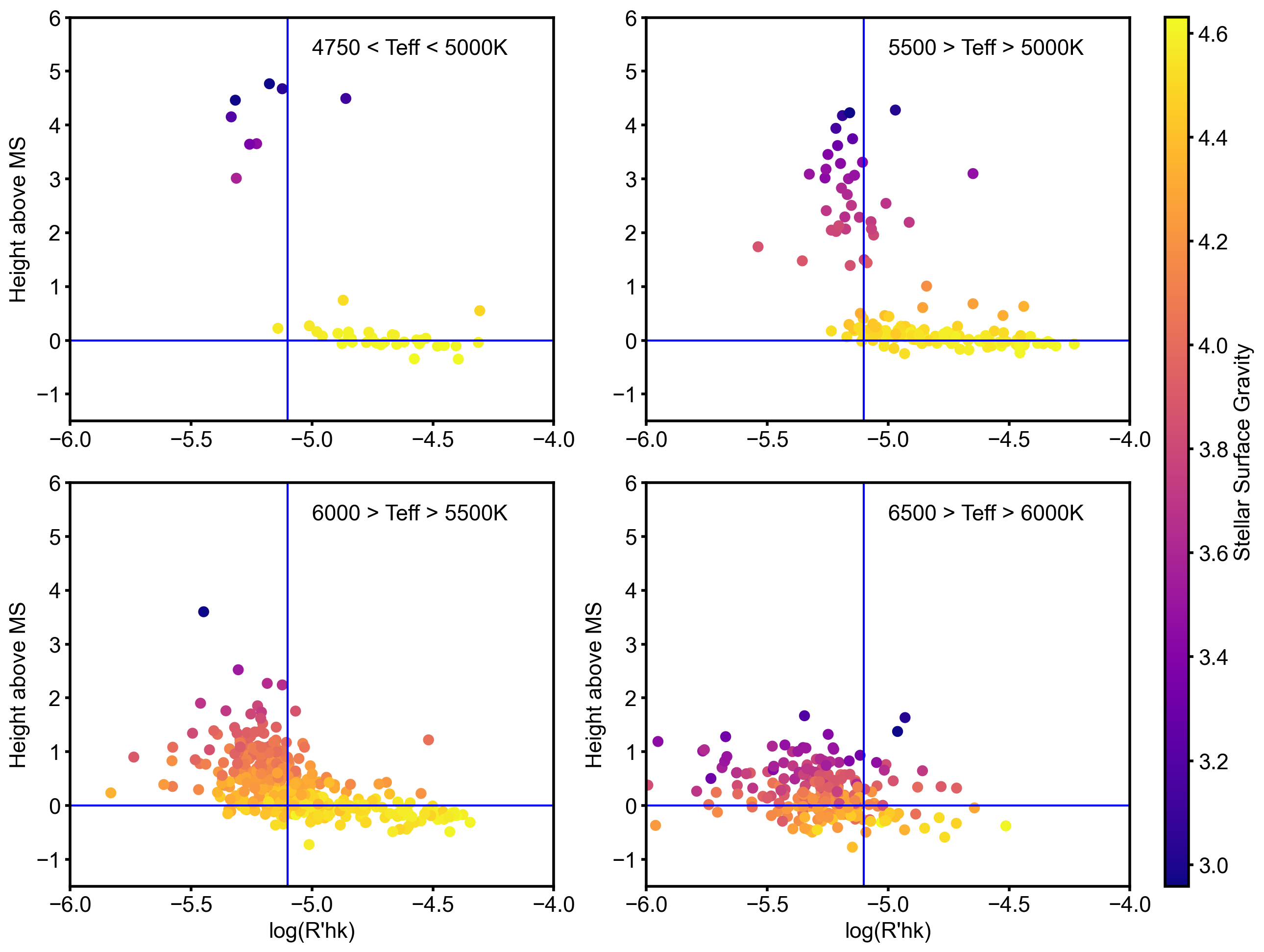}
\caption{ Examining delta-magnitude above the main-sequence \citep{Wright2004b} vs \rphk, we expect stars that are experiencing magnetic activity minima to be in the bottom left quadrant of these plots. Stars that populate the upper left quadrant tend to be sub-giants with large radii that are likely have their magnetic fields decoupled from their convective zones. The vertical line at \rphk = -5.1 is an arbitrary  but reasonable division between active and inactive stars \cite{Henry1996}. \cite{Saar2011} states that is not a good cutoff for MM star consideration because it was made before Hipparcos helped define evolutionary state of a large sample of stars. We focus on main-sequence stars with \teff < 5500 K because we expect them not to populate very inactive regime, except in extreme circumstances such as Maunder Minimum stars. } 
\label{fig:mm_search_delta_mag}
\end{figure*}  

\begin{figure*}
\includegraphics[width = 2.1\columnwidth]{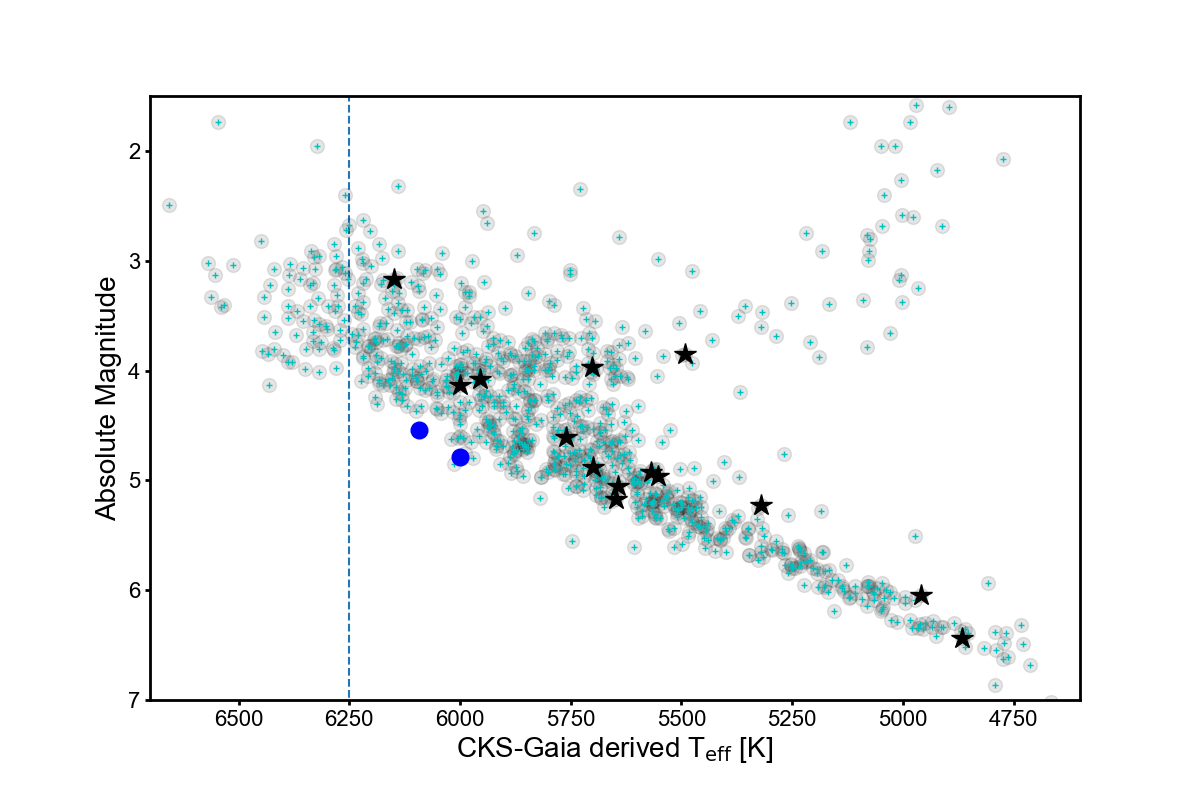}
\caption{
This color magnitude diagram highlights several sub-categories of stars and allows identification of Maunder Minimum (MM) candidate stars, following the logic of \cite{Wright2004b}. Starting the CKS-HK stellar sample as the gray background of data points. Black star symbols highlight exceptionally metal poor stars with [Fe/H]$ < -0.4$. Blue circles represent stars that are 0.4 magnitudes below the main-sequence and are 'very inactive' with \rphk $< -5.2$ as the threshold. Stars of very low stellar activity that fall below the main-sequence are candidates for MM status. The vertical line denotes the temperature of the Kraft Break, above which we do not expect stars to have an active convective zone contributing to stellar activity.  The two blue data points are KOI-4144 and KOI-5236. (KOI-1531 misses the \rphk < -5.2 cutoff). } 
\label{fig:mm_search_hr}
\end{figure*}  

\begin{figure*}
\includegraphics[width = 2.1\columnwidth]{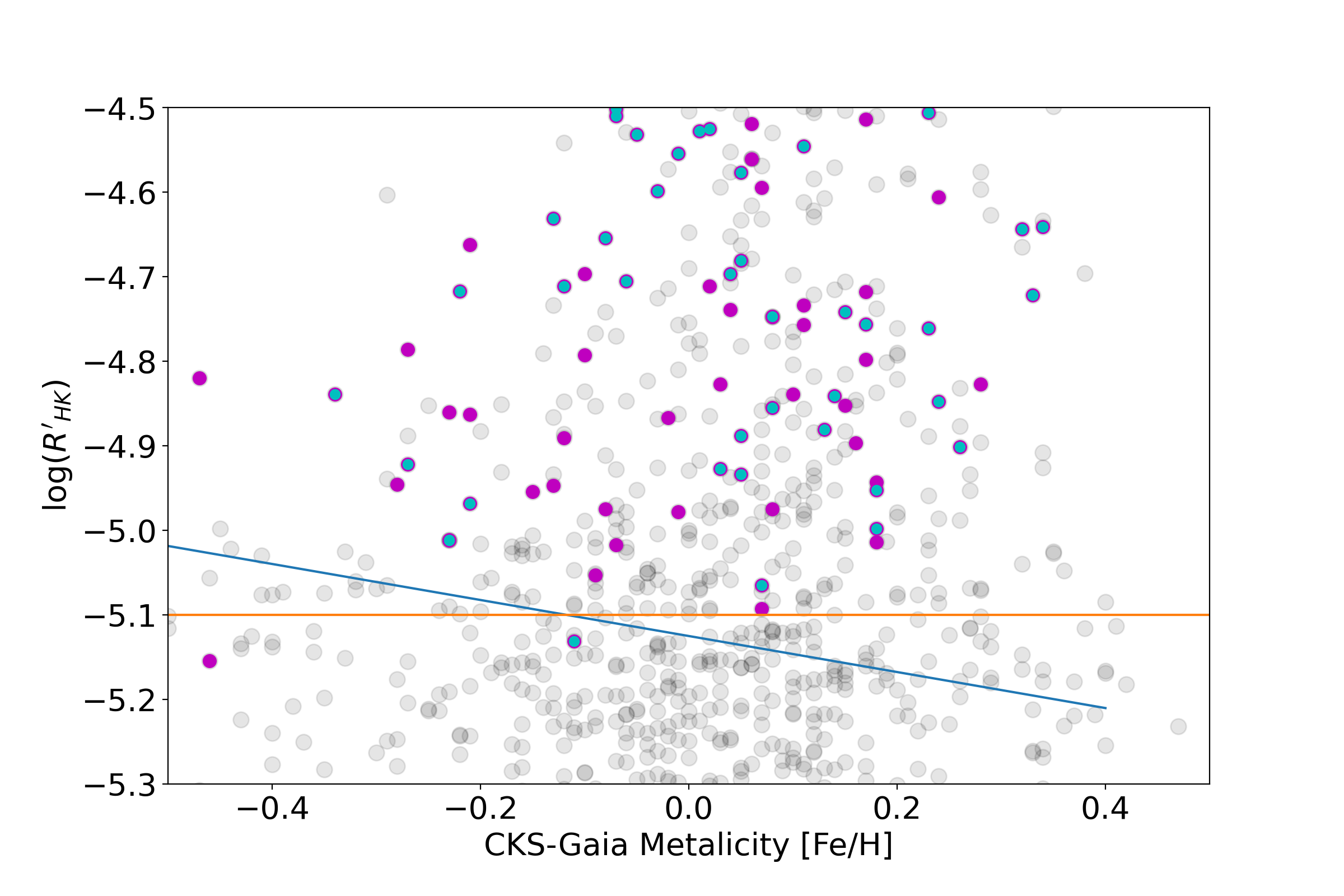}
\caption{We search for the dependence of metallicity on the lowest level of activity for dwarf stars. Inspired by \cite{Saar2011}, figure 2, we examine metallicity vs \rphk, with the blue diagonal line representing a minimum activity as a function of metallicity for dwarf stars. Our full set of 879 stars is shown in gray with magenta stars representing stars with \teff less than 5300 K and the cyan stars have delta magnitude above the main-sequence of greater than 0.5.  According to the criteria from \cite{Saar2011}, these are candidate MM stars, but knowing something about their long term variability is also required. Since we do not have that with our sample, they must remain candidates. Contrary to \cite{Saar2011}, we do not see a dependence on metallicity for low activity stars. Perhaps a sample of extremely low metallicity stars would offer a stronger case for a correlation between activity and metallicity.
} 
\label{fig:mm_search_fe}
\end{figure*}  

\begin{figure*}
\begin{center}
    
\includegraphics[width = 1.0\columnwidth]{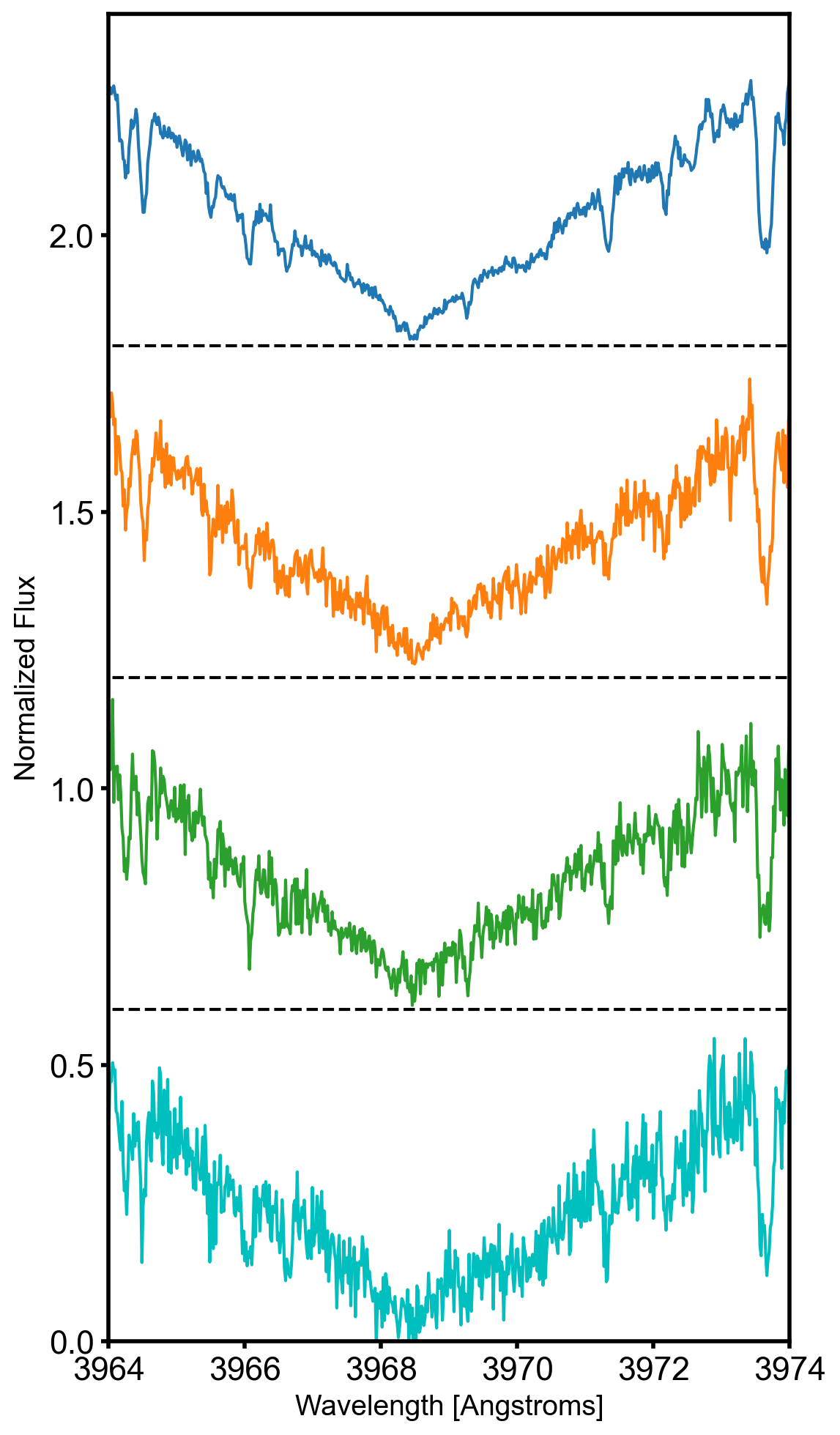}
\caption{ These four spectra represent the four most inactive stars in our survey. Each star is vertically offset with the dotted line representing zero flux. The top panel shows WASP-12 with exceptionally low activity and a spectrum with high SNR. The bottom three plots show that for stars with very little activity, it is difficult to visually distinguish between inactive and very-inactive stars. Below WASP-12 are KOI-2786, KOI-2906, and KOI-2833. Their \rphk values are -6.122, -5.994, and -5.962 respectively.} 
\end{center}
\label{fig:fig_most_inactive_spec}
\end{figure*} 

\begin{figure*}
\includegraphics[width = 1.8\columnwidth]{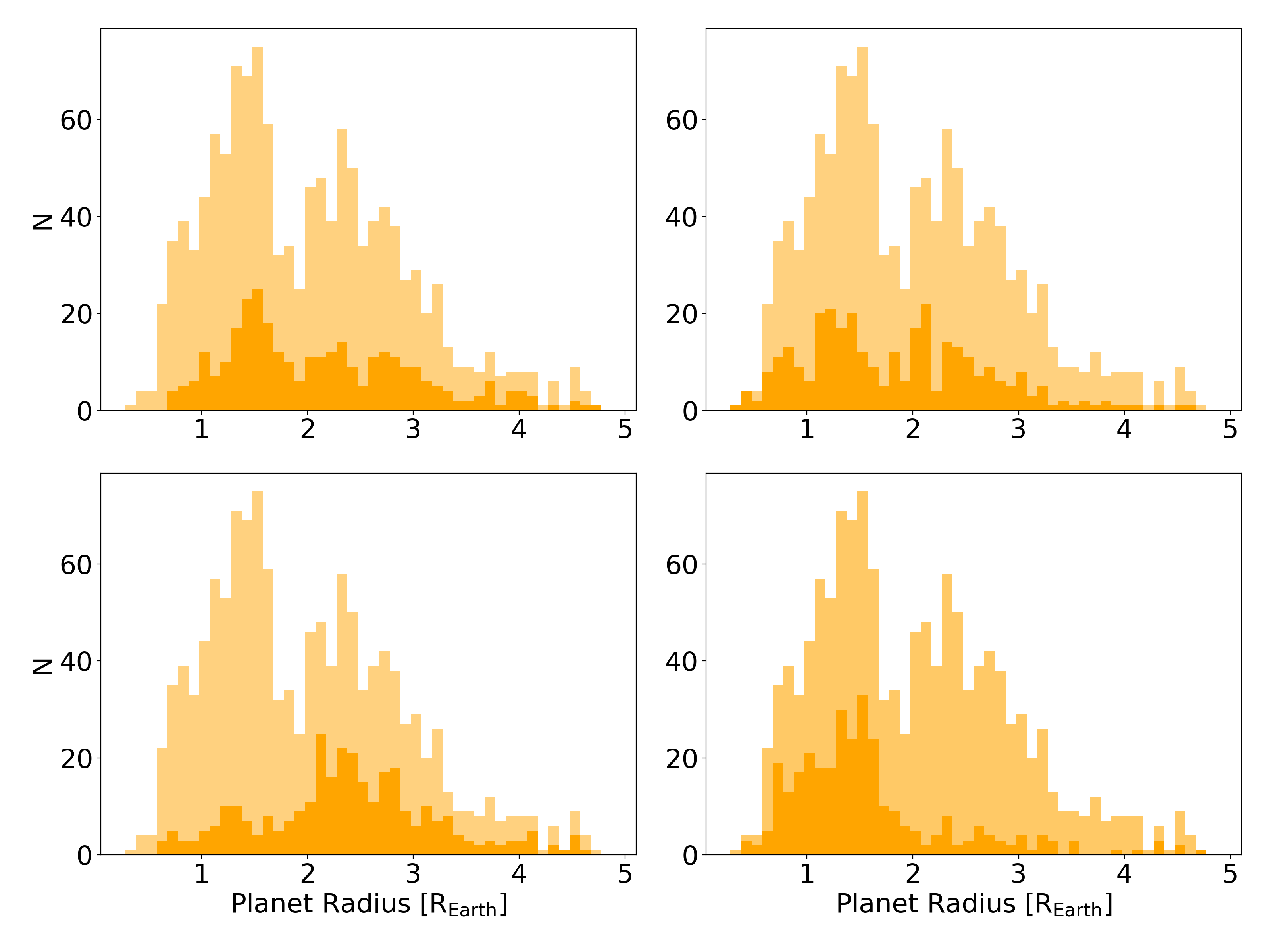}
\centering
\caption{ In the background of each plot, the CKS-HK planet population of 773 stars hosting 1243 small planets is plotted.  In the two top panels we examine the planet radius histogram to check see how the super-Earth and mini-Neptune populations correspond to the most and least chromospherically active planet-host stars. 
Top-left) As the dark color we plot the top quartile of planets in the \rphk distribution. 
Top-Right) As the dark color we plot the bottom quartile of planets in the \rphk distribution. We do not see an abundance of super-Earths in the active stars, nor an under-abundance in the sub-Neptunes as we might get if chromospheric activity was highly correlated with planet radius. The pattern present in the insolation plots is not reflected here.
Bottom-Left) The quartile of stars with the lowest insolation is plotted in dark orange showing an over-abundance of sub-Neptunes. 
Bottom-right) The quartile of planets with the highest insolation is plotted, showing an over-abundance of super-Earths. This is consistent with the results of \cite{Fulton2017}.
}
\label{fig:Rp_insolation}
\end{figure*} 

\begin{figure*}
\includegraphics[width = 2.1\columnwidth]{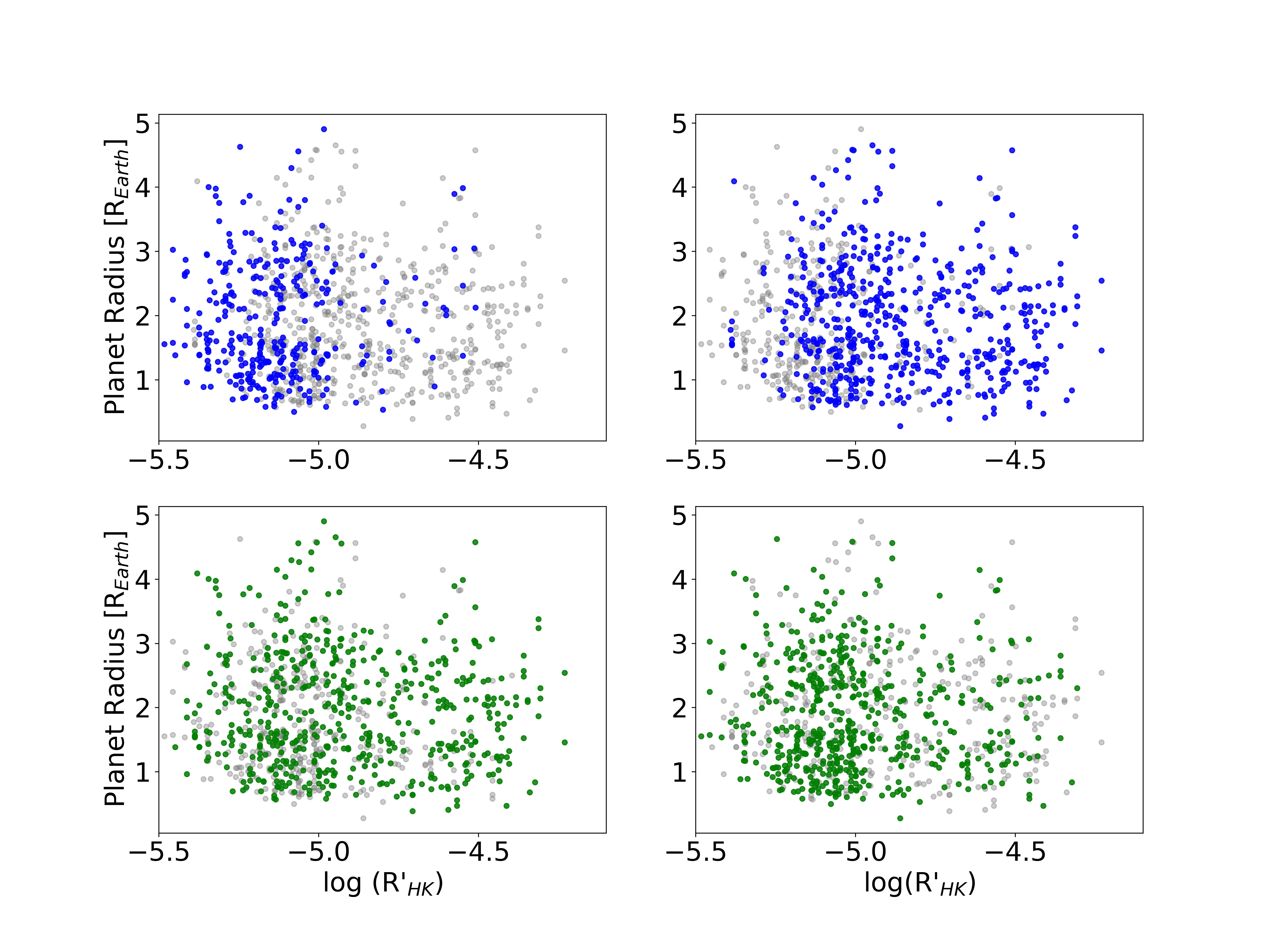}
\caption{ Planet radii as a function of chromospheric Activity for \emph{Kepler}'s small planets. The top panels show stars hotter than the sun (\teff > 5770 K, left) and those cooler than the sun, right. The gray stars show the full sample of 912 planets that pass our quality criteria.. The bottom row shows the CKS-HK sample divided at the median metallicity of 0.05 with higher metallicity on the left and lower metallicity on the right.} 
\label{fig:rphk_rp_teff}
\end{figure*} 

The reliability of the metric \rphk depends on the inputs as defined in \cite{Noyes1984}. Modern studies have created new variations on \rphk such as \citep{Hall2007} who prefer to use $F_{HK}$ and delta $F_{HK}$. This metric provides a different but related way for obtaining a minimum activity level.  Improvements in the calculation or \rphk or of the bolometric corrections could be useful in identifying the least active stars.  \cite{Mittag2013} created the log (R$^{+}_{HK}$) metric which separates the flux in the H\&K lines into the basal, photospheric and chromospheric components to create a metric at which the minimum activity level is 0 by definition.
\cite{Saar2011} requires MM stars to have a minimum activity level, but does not assume it is based only on \teff, which was assumed by \cite{Noyes1984}. This leads to a minimum level of activity that might indicate MM stars vary as a function of \vsini, \logg, \feh as well as \teff. Additionally, it requires knowing something about the longer term variation, which can be variation of S-value over time \citep{Saar2011}, or photometry \citep{Hall2007} or X-rays \citep{Judge2007}. \citep{Judge2007} claim that a MM star can only be confirmed if we know something about the X-ray flux, which can only be produced by a stellar dynamo. 

We define the height above the main-sequence using the \cite{Wright2004b} relation and search for stars that are both inactive, \rphk less than -5.1, and near or below the main-sequence.  To identify interesting candidates we focus on the stellar temperature range of 4700 K to 5500 K, where the main-sequence population and sub-giant populations are well separated. Knowing that MM stars are rare, we are looking for outliers from the populations. In Figure \ref{fig:mm_search_delta_mag}, we identify no stars in the coolest bin that are MM candidates. In the 5500 K - 6000 K bin, we see a set of stars, the most extreme of which may be an MM candidate. The best candidate in the 5500 K to 6000 K bin is KOI-1531. There is a population of stars in the lower right quadrants for the temperatures 6000 K - 6500 K. There are no outliers, but rather stars that naturally fall into the bottom left box. We expect this because near the Kraft Break at \teff of 6250 K, stars become fully radiative, and the sort of activity that we are probing is not viable through the same physical mechanisms as for cooler stars.

We highlight cool stars that lie far below the main-sequence in a Temperature - Magnitude diagram (Figure \ref{fig:mm_search_hr}). There are three stars that are more than 0.4 magnitude below the main-sequence, as defined by \cite{Wright2004b} and are cooler than 6250K, the Kraft Break. They are KOI-1531, KOI-4144 and KOI-5236. KOI-5236 has a planet with a  period greater than 500d, and maybe false positive planet candidate according to Community Follow-up Program (CFOP) \cite{Gautier2010}. It has \teff = 6100 K, \logg = 4.41 and \rphk = -5.365. It is very inactive and has a 1.9" companion, but no detection of a secondary star in the spectrum.  The nearby star makes the likelihood of the star being an MM star less certain.  KOI-4144 has \teff of 6000 K, \logg = 4.49, especially large RMS3 value from \emph{Kepler} \citep{Christiansen2012}, distance of 860 pc. 
KOI-1531 is the furthest below the main-sequence and it has a near sun-like \teff , and \logg with \rphk = -5.012. It has a photometric rotation period from \cite{Angus2018}  of 23 days and an activity rotation period of 34 days. This 30\% discrepancy mismatch between the photometric and activity rotation periods is supportive of the theory of WMB. If this is the case, the star is old and inactive, but still has the rotation period of a younger star. 

\cite{Henry1996} states that stars with \rphk $\leq$ -5.1 are in MM states. \cite{Wright2004b} claims that without knowing \logg or height above the main-sequence, it is easy to confuse subgiants with MM stars. This was problematic both because subgiants have inherently lower activity, so they can fall in the same parameter space as MM stars, and because the search for stars in states of magnetic minima are more compelling for sun-like stars. If the conclusions from \cite{Wright2004b} are correct, 
we expect stars with \rphk < -5.1 to be evolved stars. The bottom row of Figure \ref{fig:hist-9panel} shows stars with smaller radii are more active, populating the Inactive and Active bins.
Our three best candidates above are distinctly not subgiants. We have an unusually large number of stars relative to the nearby star catalogs, with a full population of stars with \rphk $<$ -5.1.

\cite{Hall2007} studied eighteen solar type stars and concluded that \rphk is not a good metric to identify stars in a magnetic minimum state. This is due to to the unknown effects of metallicity and stellar surface gravity. They claim that a better metric is delta F$_{HK}$, which is the measure of the flux caused by magnetic activity in the H and K lines. It is robust against high stellar rotation as well. This agrees with \cite{Saar2006}, which noted that the minimum activity of a star is dependent on metallicity. Specifically, metal-poor dwarfs have a higher minimum \rphk than dwarfs of solar metallicity.

X-rays play an important role in observing and assessing chromospheric activity and searching for MM stars because it is difficult to explain the creation of X-rays without the magnetic field processes associated with chromospheric activity detected in the optical. We reserve the incorporation of X-ray data for future work.

\cite{Wright2004b} had precise parallaxes from Hipparcos but lacked precise metallicity and stellar surface gravity required to separate their relationship to stellar activity.  We now have excellent metallicities, so we can check if stars that are below the main-sequence are very metal poor. \cite{Saar2011} provided a functional cutoff of in the plotting of \rphk  as a function of metallicity.

Using a similar analysis to \cite{Saar2011}, we can plot chromospheric activity as function of stellar metallicity, limit our search to dwarf stars and identify the least active stars.  Broadly speaking, we do not see the trend of minimum \rphk that is noted by \cite{Saar2011}. While we lack many stars beyond +/- 0.4 dex that give the most leverage in assessing effects of metallicity.  Figure \ref{fig:mm_search_fe} shows no visible trend in metallicity for stars in the CKS-HK sample, although with limited parameter space. Specifically, we find KOI-241 and KOI-2498 are the two least active stars, and confirm, by eye, that they have high quality S-values. They are both quite cool stars compared to the solar temperature inactive stars that were previously discussed, with \teff of 4960 K and 5128 K, respectively. Their metallicities are -0.46 and -0.11.  The \citep{David2021} rotation periods are 32.7 d and 15.2, respectively, but both are listed as ambiguous or lacking a clear cycle.  An additional test would be to check the stars’ variation over time. With only single epoch measurements, we can only say these stars are currently in low activity states, but not necessarily in extended MM states. 

\subsubsection{Activity, Close in Planets, and Toroidal Gas Rings}

The work of \cite{Fossati2013} revealed that the star system hosting ultra-short period hot Jupiter planet WASP-12b shows evidence of a gas ring surrounding the star which is due to excessively low flux in the Mg II H and K lines, located at 2586 \AA. The \cahk  lines confirm this interpretation along with measurements of the strongly varying near UV flux. We searched for similar occurrences of incredibly low \rphk values around similar type planet hosting systems and find none as compelling as WASP-12 which has a \rphk value of -5.50 and \teff  6250 K.

We searched our sample for stars with a \teff  between 5800 and 6250K, planet orbital periods less than 1 day and stellar surface gravity greater than 4.2, bracketing the properties of the WASP-12 system. Of the two systems that have planet radii larger than ten Earth-radii, both are eclipsing binaries from CFOP \citep{Gautier2010}. For planets with radii less than ten Earth-radii and the same stellar property limitations as previously listed, ten systems remain, the largest with a planet radius of three Earth-radii. Three stars have \rphk less than -5.2.
KOI-2717 has a \rphk value = -5.226 and shows visible evidence of emission below the solar level, in a high SNR spectrum. KOI-4072 (\rphk = -5.241) and KOI-4144 (\rphk = -5.306) show modest evidence of very low flux, but the SNR makes visual confirmation difficult. Reobservation of these three stars at higher SNR could help confirm their activity level. Figure \ref{fig:fig_most_inactive_spec} shows WASP-12 \cahk spectral features along with the three most inactive stars in our survey. We do not find convincing evidence of a toroidal gas ring, although our survey does not contain many hot Jupiters. Transit surveys such as TESS will detect many transiting bright stars enabling a larger survey of stars to search toroidal gas rings.

\emph{Kepler} stars are more metal-rich than the solar neighborhood so the paucity of hot Jupiters is not related to metallicity in a similar way to the hot Jupiter occurrence 
in the solar neighborhood \citep{Petigura2018}. The bias of the \emph{Kepler} sample also contains a large number of sub-giants with have inherently lower chromospheric emission. Using a sample of sun-like stars hosting hot Jupiters would provide a more complete search space than the CKS sample of transiting planet hosts. Using a large sample of hot Jupiter host star spectra to search for such anomalous emission is more promising than using the CKS-HK data set.

Future work on this could include examination of the Rossiter McLaughlin sequences of inflated hot Jupiters to search for variations in the S-values over the course of the transit. Similarly, if a large sample of hot Jupiters were sampled such as those found by TESS, perhaps more stars similar to WASP-12 would be found. 

\subsection{Planet Radii and Chromospheric Activity}
\label{sec:planets_activity}

Identification of the small planet radius gap (Fulton gap) at roughly 1.8 $R_{Earth}$ both theoretically and observationally has been of great interest to planet formation theory. Stellar insolation at the planets' average orbital distance is strongly correlated with planet radius on the small side of the radius gap receiving higher insolation and planets on the larger radius side receiving less. Stellar mass plays a critical role in shaping the exact radius at which the gap falls \citep{Petigura2022} and predictions of the time frame over which the gap is sculpted ranges from less than 1 Gyr \citep{Berger2020b} to several Gyr timescales \citep{david2022}.

Acknowledging that stars become less active as they age, making age and activity degenerate, we examine the relationship between activity and planet radius directly with our CKS-HK planet sample. We plot histograms of the most active and least active stars and compare those planet populations to the planets with the most and least insolation flux.  Figure \ref{fig:Rp_insolation} shows four histograms of the CKS-HK planet sample. Each panel shows the full sample histogram with a different sub-set overplotted. In the top two panels, we see the least active quartile overplotted (left) and the most active quartile(right). The bottom panels show the planets receiving the lowest quartile (left) and the highest quartile of insolation flux (right). Without absolute occurrence of planet in each bin as was shown in \cite{Fulton2018}, we can still identify the previously established trend of small mass planets receiving higher insolation. In the top panels we  see no obvious differences in the two quartiles that separate planet hosts in the most and least active quartiles.   We expect the super-Earths to receive the most insolation and mini-Neptunes to receive less. 
If the present chromospheric activity of a star is correlated to the planet radius, we would expect a similar pattern in the top panels. Instead, we see a peak of super-earths in the most chromospherically quiet stars. We also searched for trends and correlations in the \rphk vs orbital separation parameter space and found no significant trends.

The connection between planet size and chromospheric activity is degenerate with age, complicating the interpretation. We plot the chromospheric activity vs planet radius in Figure \ref{fig:rphk_rp_teff}, subdividing the sample by both \teff and \feh. For a full explanation of isochronal age dependence on the radius gap see \cite{Berger2020b, david2022}.

\section{Conclusion}

We present chromospheric activity measurements of the well-studied CKS-Gaia sample of transiting planet host stars via the \cahk spectral features which are known to track magnetic activity in solar type stars. These novel measurements are used along with the fundamental stellar properties including temperature, surface gravity and metallicity and photometric rotation periods to explore fundamental relationships in the age-rotation-activity regime of stellar astrophysics.

We also have used the chromospheric activity measurements to check for a dependency of metallicity on the rotation periods measured from photometry and the stars’ corresponding Rossby numbers. Using rotation periods determined from Kepler light curves, we have searched for stellar over-abundances among long and short rotation periods in the temperature-period plane. With our single epoch spectra, we have identified very inactive stars, but we do not yet have the multi-year time series needed to verify stars in Maunder Minimum (aka Magnetic Minimum) like states. 

From the large sample of precise stellar properties and activity results presented in this paper, we find no significant evidence for metallicity dependent activity relations within the metallicity range of -0.2 to +0.3. Our results are supportive of the theory of weakened magnetic braking of stellar spin-down in the form of discrepancies between activity determined and photometrically determined rotation periods. While 1 Gyr previously has been suggested as critical age juncture, we find no significant evidence for a change in the activity-period relationship at this age using chromospherically derived ages. The activity-period relationship presented here, along with recently discovered nuances discovered in the temperature-period plane such as the Rossby Ridge and their relative independence from stellar metallicity, can all inform future theoretical studies of stellar rotation-activity relationships, and understanding the physics of the underlying stellar magnetic dynamos.

\emph{Acknowledgments}
We would like to thank Travis Metcalfe and Luke Bouma for helpful discussions on early versions of the paper and Ryan Rubenzahl, Sarah Blunt and Aida Behmard for feedback during group discussions.

Some of the data presented herein were obtained at the W. M. Keck Observatory, which is operated as a scientific partnership among the California Institute of Technology, the University of California and the National Aeronautics and Space Administration. The Observatory was made possible by the generous financial support of the W. M. Keck Foundation. This research has made use of NASA's Astrophysics
Data System.
L.M.W. acknowledges support from the NASA-Keck Key Strategic Mission Support program (grant no. 80NSSC19K1475) and the NASA Exoplanet Research Program (grant no. 80NSSC23K0269).
Support was provided by the Simons Foundation grant "Planetary Context of Habitability and Exobiology”

The authors wish to recognize and acknowledge the very significant cultural role and reverence that the summit of Maunakea has always had within the indigenous Hawaiian community.  We are most fortunate to have the opportunity to conduct observations from this mountain. 

This project would not have been possible without major allocations of Keck telescope time from the University of California, California Institute of Technology, the University of Hawaii and NASA. This work utilized the Exoplanet Archive, SIMBAD, the Community Follow-up Program, and data from NASA's \emph{Kepler} Space Telescope.

\facility{Keck:I (HIRES)}, \emph{Kepler} 

\software{We made use of the following publicly available Python modules: astropy (Astropy Collaboration et al. 2013),
 matplotlib (Hunter 2007), numpy/scipy (van der Walt et al. 2011), and pandas
(McKinney 2010). Interactive Data Language (IDL) was used to extract the spectral line information. }
          
\bibliography{main}{}

\begin{thebibliography}{}
\expandafter\ifx\csname natexlab\endcsname\relax\def\natexlab#1{#1}\fi
\providecommand{\url}[1]{\href{#1}{#1}}
\providecommand{\dodoi}[1]{doi:~\href{http://doi.org/#1}{\nolinkurl{#1}}}
\providecommand{\doeprint}[1]{\href{http://ascl.net/#1}{\nolinkurl{http://ascl.net/#1}}}
\providecommand{\doarXiv}[1]{\href{https://arxiv.org/abs/#1}{\nolinkurl{https://arxiv.org/abs/#1}}}

\bibitem[{{Aigrain} {et~al.}(2012){Aigrain}, {Pont}, \& {Zucker}}]{Aigrain2012}
{Aigrain}, S., {Pont}, F., \& {Zucker}, S. 2012, \mnras, 419, 3147,
  \dodoi{10.1111/j.1365-2966.2011.19960.x}

\bibitem[{{Akana Murphy} {et~al.}(2021){Akana Murphy}, {Kosiarek}, {Batalha},
  {Gonzales}, {Isaacson}, {Petigura}, {Weiss}, {Grunblatt}, {Ciardi}, {Fulton},
  {Hirsch}, {Behmard}, \& {Rosenthal}}]{Murphy2021}
{Akana Murphy}, J.~M., {Kosiarek}, M.~R., {Batalha}, N.~M., {et~al.} 2021, \aj,
  162, 294, \dodoi{10.3847/1538-3881/ac2830}

\bibitem[{{Amard} \& {Matt}(2020)}]{Amard2020A}
{Amard}, L., \& {Matt}, S.~P. 2020, \apj, 889, 108,
  \dodoi{10.3847/1538-4357/ab6173}

\bibitem[{{Angus} {et~al.}(2018){Angus}, {Morton}, {Aigrain}, {Foreman-Mackey},
  \& {Rajpaul}}]{Angus2018}
{Angus}, R., {Morton}, T., {Aigrain}, S., {Foreman-Mackey}, D., \& {Rajpaul},
  V. 2018, \mnras, 474, 2094, \dodoi{10.1093/mnras/stx2109}

\bibitem[{{Avallone} {et~al.}(2022){Avallone}, {Tayar}, {van Saders}, {Berger},
  {Claytor}, {Beaton}, {Teske}, {Godoy-Rivera}, \& {Pan}}]{Avallone2022}
{Avallone}, E.~A., {Tayar}, J.~N., {van Saders}, J.~L., {et~al.} 2022, \apj,
  930, 7, \dodoi{10.3847/1538-4357/ac60a1}

\bibitem[{{Baliunas} {et~al.}(1995){Baliunas}, {Donahue}, {Soon}, {Horne},
  {Frazer}, {Woodard-Eklund}, {Bradford}, {Rao}, {Wilson}, {Zhang}, {Bennett},
  {Briggs}, {Carroll}, {Duncan}, {Figueroa}, {Lanning}, {Misch}, {Mueller},
  {Noyes}, {Poppe}, {Porter}, {Robinson}, {Russell}, {Shelton}, {Soyumer},
  {Vaughan}, \& {Whitney}}]{Baliunas1995}
{Baliunas}, S.~L., {Donahue}, R.~A., {Soon}, W.~H., {et~al.} 1995, \apj, 438,
  269, \dodoi{10.1086/175072}

\bibitem[{{Batalha} {et~al.}(2011){Batalha}, {Borucki}, {Bryson}, {Buchhave},
  {Caldwell}, {Christensen-Dalsgaard}, {Ciardi}, {Dunham}, {Fressin},
  {Gautier}, {Gilliland}, {Haas}, {Howell}, {Jenkins}, {Kjeldsen}, {Koch},
  {Latham}, {Lissauer}, {Marcy}, {Rowe}, {Sasselov}, {Seager}, {Steffen},
  {Torres}, {Basri}, {Brown}, {Charbonneau}, {Christiansen}, {Clarke},
  {Cochran}, {Dupree}, {Fabrycky}, {Fischer}, {Ford}, {Fortney}, {Girouard},
  {Holman}, {Johnson}, {Isaacson}, {Klaus}, {Machalek}, {Moorehead},
  {Morehead}, {Ragozzine}, {Tenenbaum}, {Twicken}, {Quinn}, {VanCleve},
  {Walkowicz}, {Welsh}, {Devore}, \& {Gould}}]{Batalha2011}
{Batalha}, N.~M., {Borucki}, W.~J., {Bryson}, S.~T., {et~al.} 2011, \apj, 729,
  27, \dodoi{10.1088/0004-637X/729/1/27}

\bibitem[{{Batalha} {et~al.}(2013){Batalha}, {Rowe}, {Bryson}, {Barclay},
  {Burke}, {Caldwell}, {Christiansen}, {Mullally}, {Thompson}, {Brown},
  {Dupree}, {Fabrycky}, {Ford}, {Fortney}, {Gilliland}, {Isaacson}, {Latham},
  {Marcy}, {Quinn}, {Ragozzine}, {Shporer}, {Borucki}, {Ciardi}, {Gautier},
  {Haas}, {Jenkins}, {Koch}, {Lissauer}, {Rapin}, {Basri}, {Boss}, {Buchhave},
  {Carter}, {Charbonneau}, {Christensen-Dalsgaard}, {Clarke}, {Cochran},
  {Demory}, {Desert}, {Devore}, {Doyle}, {Esquerdo}, {Everett}, {Fressin},
  {Geary}, {Girouard}, {Gould}, {Hall}, {Holman}, {Howard}, {Howell},
  {Ibrahim}, {Kinemuchi}, {Kjeldsen}, {Klaus}, {Li}, {Lucas}, {Meibom},
  {Morris}, {Pr{\v{s}}a}, {Quintana}, {Sanderfer}, {Sasselov}, {Seader},
  {Smith}, {Steffen}, {Still}, {Stumpe}, {Tarter}, {Tenenbaum}, {Torres},
  {Twicken}, {Uddin}, {Van Cleve}, {Walkowicz}, \& {Welsh}}]{Batalha2013}
{Batalha}, N.~M., {Rowe}, J.~F., {Bryson}, S.~T., {et~al.} 2013, \apjs, 204,
  24, \dodoi{10.1088/0067-0049/204/2/24}

\bibitem[{{Baum} {et~al.}(2022){Baum}, {Wright}, {Luhn}, \&
  {Isaacson}}]{Baum2022}
{Baum}, A.~C., {Wright}, J.~T., {Luhn}, J.~K., \& {Isaacson}, H. 2022, \aj,
  163, 183, \dodoi{10.3847/1538-3881/ac5683}

\bibitem[{{Berger} {et~al.}(2020{\natexlab{a}}){Berger}, {Huber}, {Gaidos},
  {van Saders}, \& {Weiss}}]{Berger2020b}
{Berger}, T.~A., {Huber}, D., {Gaidos}, E., {van Saders}, J.~L., \& {Weiss},
  L.~M. 2020{\natexlab{a}}, \aj, 160, 108, \dodoi{10.3847/1538-3881/aba18a}

\bibitem[{{Berger} {et~al.}(2020{\natexlab{b}}){Berger}, {Huber}, {van Saders},
  {Gaidos}, {Tayar}, \& {Kraus}}]{Berger2020a}
{Berger}, T.~A., {Huber}, D., {van Saders}, J.~L., {et~al.} 2020{\natexlab{b}},
  \aj, 159, 280, \dodoi{10.3847/1538-3881/159/6/280}

\bibitem[{{Blunt} {et~al.}(2023){Blunt}, {Carvalho}, {David}, {Beichman},
  {Zink}, {Gaidos}, {Behmard}, {Bouma}, {Cody}, {Dai}, {Foreman-Mackey},
  {Grunblatt}, {Howard}, {Kosiarek}, {Knutson}, {Rubenzahl}, {Beard},
  {Chontos}, {Giacalone}, {Hirano}, {Johnson}, {Lubin}, {Akana Murphy},
  {Petigura}, {Van Zandt}, \& {Weiss}}]{Blunt2023}
{Blunt}, S., {Carvalho}, A., {David}, T.~J., {et~al.} 2023, \aj, 166, 62,
  \dodoi{10.3847/1538-3881/acde78}

\bibitem[{{Borucki} {et~al.}(2010){Borucki}, {Koch}, {Brown}, {Basri},
  {Batalha}, {Caldwell}, {Cochran}, {Dunham}, {Gautier}, {Geary}, {Gilliland},
  {Howell}, {Jenkins}, {Latham}, {Lissauer}, {Marcy}, {Monet}, {Rowe}, \&
  {Sasselov}}]{Borucki2010}
{Borucki}, W.~J., {Koch}, D.~G., {Brown}, T.~M., {et~al.} 2010, \apjl, 713,
  L126, \dodoi{10.1088/2041-8205/713/2/L126}

\bibitem[{{Borucki} {et~al.}(2011{\natexlab{a}}){Borucki}, {Koch}, {Basri},
  {Batalha}, {Boss}, {Brown}, {Caldwell}, {Christensen-Dalsgaard}, {Cochran},
  {DeVore}, {Dunham}, {Dupree}, {Gautier}, {Geary}, {Gilliland}, {Gould},
  {Howell}, {Jenkins}, {Kjeldsen}, {Latham}, {Lissauer}, {Marcy}, {Monet},
  {Sasselov}, {Tarter}, {Charbonneau}, {Doyle}, {Ford}, {Fortney}, {Holman},
  {Seager}, {Steffen}, {Welsh}, {Allen}, {Bryson}, {Buchhave},
  {Chandrasekaran}, {Christiansen}, {Ciardi}, {Clarke}, {Dotson}, {Endl},
  {Fischer}, {Fressin}, {Haas}, {Horch}, {Howard}, {Isaacson}, {Kolodziejczak},
  {Li}, {MacQueen}, {Meibom}, {Prsa}, {Quintana}, {Rowe}, {Sherry},
  {Tenenbaum}, {Torres}, {Twicken}, {Van Cleve}, {Walkowicz}, \&
  {Wu}}]{Borucki2011}
{Borucki}, W.~J., {Koch}, D.~G., {Basri}, G., {et~al.} 2011{\natexlab{a}},
  \apj, 728, 117, \dodoi{10.1088/0004-637X/728/2/117}

\bibitem[{{Borucki} {et~al.}(2011{\natexlab{b}}){Borucki}, {Koch}, {Basri},
  {Batalha}, {Brown}, {Bryson}, {Caldwell}, {Christensen-Dalsgaard}, {Cochran},
  {DeVore}, {Dunham}, {Gautier}, {Geary}, {Gilliland}, {Gould}, {Howell},
  {Jenkins}, {Latham}, {Lissauer}, {Marcy}, {Rowe}, {Sasselov}, {Boss},
  {Charbonneau}, {Ciardi}, {Doyle}, {Dupree}, {Ford}, {Fortney}, {Holman},
  {Seager}, {Steffen}, {Tarter}, {Welsh}, {Allen}, {Buchhave}, {Christiansen},
  {Clarke}, {Das}, {D{\'e}sert}, {Endl}, {Fabrycky}, {Fressin}, {Haas},
  {Horch}, {Howard}, {Isaacson}, {Kjeldsen}, {Kolodziejczak}, {Kulesa}, {Li},
  {Lucas}, {Machalek}, {McCarthy}, {MacQueen}, {Meibom}, {Miquel}, {Prsa},
  {Quinn}, {Quintana}, {Ragozzine}, {Sherry}, {Shporer}, {Tenenbaum}, {Torres},
  {Twicken}, {Van Cleve}, {Walkowicz}, {Witteborn}, \& {Still}}]{Borucki2011b}
---. 2011{\natexlab{b}}, \apj, 736, 19, \dodoi{10.1088/0004-637X/736/1/19}

\bibitem[{{Borucki} {et~al.}(2013){Borucki}, {Agol}, {Fressin}, {Kaltenegger},
  {Rowe}, {Isaacson}, {Fischer}, {Batalha}, {Lissauer}, {Marcy}, {Fabrycky},
  {D{\'e}sert}, {Bryson}, {Barclay}, {Bastien}, {Boss}, {Brugamyer},
  {Buchhave}, {Burke}, {Caldwell}, {Carter}, {Charbonneau}, {Crepp},
  {Christensen-Dalsgaard}, {Christiansen}, {Ciardi}, {Cochran}, {DeVore},
  {Doyle}, {Dupree}, {Endl}, {Everett}, {Ford}, {Fortney}, {Gautier}, {Geary},
  {Gould}, {Haas}, {Henze}, {Howard}, {Howell}, {Huber}, {Jenkins}, {Kjeldsen},
  {Kolbl}, {Kolodziejczak}, {Latham}, {Lee}, {Lopez}, {Mullally}, {Orosz},
  {Prsa}, {Quintana}, {Sanchis-Ojeda}, {Sasselov}, {Seader}, {Shporer},
  {Steffen}, {Still}, {Tenenbaum}, {Thompson}, {Torres}, {Twicken}, {Welsh}, \&
  {Winn}}]{Borucki2013}
{Borucki}, W.~J., {Agol}, E., {Fressin}, F., {et~al.} 2013, Science, 340, 587,
  \dodoi{10.1126/science.1234702}

\bibitem[{{Brandenburg} {et~al.}(2017){Brandenburg}, {Mathur}, \&
  {Metcalfe}}]{Brandenburg2017}
{Brandenburg}, A., {Mathur}, S., \& {Metcalfe}, T.~S. 2017, \apj, 845, 79,
  \dodoi{10.3847/1538-4357/aa7cfa}

\bibitem[{{Chen} \& {Rogers}(2016)}]{Chen2016}
{Chen}, H., \& {Rogers}, L.~A. 2016, \apj, 831, 180,
  \dodoi{10.3847/0004-637X/831/2/180}

\bibitem[{{Christiansen} {et~al.}(2012){Christiansen}, {Jenkins}, {Caldwell},
  {Burke}, {Tenenbaum}, {Seader}, {Thompson}, {Barclay}, {Clarke}, {Li},
  {Smith}, {Stumpe}, {Twicken}, \& {Van Cleve}}]{Christiansen2012}
{Christiansen}, J.~L., {Jenkins}, J.~M., {Caldwell}, D.~A., {et~al.} 2012,
  \pasp, 124, 1279, \dodoi{10.1086/668847}

\bibitem[{{Curtis} {et~al.}(2020){Curtis}, {Ag{\"u}eros}, {Matt}, {Covey},
  {Douglas}, {Angus}, {Saar}, {Cody}, {Vanderburg}, {Law}, {Kraus}, {Latham},
  {Baranec}, {Riddle}, {Ziegler}, {Lund}, {Torres}, {Meibom}, {Aguirre}, \&
  {Wright}}]{Curtis2020}
{Curtis}, J.~L., {Ag{\"u}eros}, M.~A., {Matt}, S.~P., {et~al.} 2020, \apj, 904,
  140, \dodoi{10.3847/1538-4357/abbf58}

\bibitem[{{Dai} {et~al.}(2020){Dai}, {Winn}, {Schlaufman}, {Wang}, {Weiss},
  {Petigura}, {Howard}, \& {Fang}}]{Dai2020}
{Dai}, F., {Winn}, J.~N., {Schlaufman}, K., {et~al.} 2020, \aj, 159, 247,
  \dodoi{10.3847/1538-3881/ab88b8}

\bibitem[{{David} {et~al.}(2022){David}, {Angus}, {Curtis}, {van Saders},
  {Colman}, {Contardo}, {Lu}, \& {Zinn}}]{david2022}
{David}, T.~J., {Angus}, R., {Curtis}, J.~L., {et~al.} 2022, \apj, 933, 114,
  \dodoi{10.3847/1538-4357/ac6dd3}

\bibitem[{{David} {et~al.}(2021){David}, {Contardo}, {Sandoval}, {Angus}, {Lu},
  {Bedell}, {Curtis}, {Foreman-Mackey}, {Fulton}, {Grunblatt}, \&
  {Petigura}}]{David2021}
{David}, T.~J., {Contardo}, G., {Sandoval}, A., {et~al.} 2021, \aj, 161, 265,
  \dodoi{10.3847/1538-3881/abf439}

\bibitem[{{Duncan} {et~al.}(1991){Duncan}, {Vaughan}, {Wilson}, {Preston},
  {Frazer}, {Lanning}, {Misch}, {Mueller}, {Soyumer}, {Woodard}, {Baliunas},
  {Noyes}, {Hartmann}, {Porter}, {Zwaan}, {Middelkoop}, {Rutten}, \&
  {Mihalas}}]{Duncan1991}
{Duncan}, D.~K., {Vaughan}, A.~H., {Wilson}, O.~C., {et~al.} 1991, \apjs, 76,
  383, \dodoi{10.1086/191572}

\bibitem[{{Eddy}(1976)}]{Eddy1976}
{Eddy}, J.~A. 1976, Science, 192, 1189, \dodoi{10.1126/science.192.4245.1189}

\bibitem[{{Fossati} {et~al.}(2013){Fossati}, {Ayres}, {Haswell}, {Bohlender},
  {Kochukhov}, \& {Fl{\"o}er}}]{Fossati2013}
{Fossati}, L., {Ayres}, T.~R., {Haswell}, C.~A., {et~al.} 2013, \apjl, 766,
  L20, \dodoi{10.1088/2041-8205/766/2/L20}

\bibitem[{{Fulton} \& {Petigura}(2018)}]{Fulton2018}
{Fulton}, B.~J., \& {Petigura}, E.~A. 2018, \aj, 156, 264,
  \dodoi{10.3847/1538-3881/aae828}

\bibitem[{{Fulton} {et~al.}(2015){Fulton}, {Weiss}, {Sinukoff}, {Isaacson},
  {Howard}, {Marcy}, {Henry}, {Holden}, \& {Kibrick}}]{Fulton2015}
{Fulton}, B.~J., {Weiss}, L.~M., {Sinukoff}, E., {et~al.} 2015, \apj, 805, 175,
  \dodoi{10.1088/0004-637X/805/2/175}

\bibitem[{{Fulton} {et~al.}(2017){Fulton}, {Petigura}, {Howard}, {Isaacson},
  {Marcy}, {Cargile}, {Hebb}, {Weiss}, {Johnson}, {Morton}, {Sinukoff},
  {Crossfield}, \& {Hirsch}}]{Fulton2017}
{Fulton}, B.~J., {Petigura}, E.~A., {Howard}, A.~W., {et~al.} 2017, \aj, 154,
  109, \dodoi{10.3847/1538-3881/aa80eb}

\bibitem[{{Gautier} {et~al.}(2010){Gautier}, {Batalha}, {Borucki}, {Cochran},
  {Dunham}, {Howell}, {Koch}, {Latham}, {Marcy}, {Buchhave}, {Ciardi}, {Endl},
  {Furesz}, {Isaacson}, {MacQueen}, {Mandushev}, \& {Walkowicz}}]{Gautier2010}
{Gautier}, Thomas~N., I., {Batalha}, N.~M., {Borucki}, W.~J., {et~al.} 2010,
  arXiv e-prints, arXiv:1001.0352.
\newblock \doarXiv{1001.0352}

\bibitem[{{Gomes da Silva} {et~al.}(2021){Gomes da Silva}, {Santos},
  {Adibekyan}, {Sousa}, {Campante}, {Figueira}, {Bossini}, {Delgado-Mena},
  {Monteiro}, {de Laverny}, {Recio-Blanco}, \& {Lovis}}]{GomesdaSilva2021}
{Gomes da Silva}, J., {Santos}, N.~C., {Adibekyan}, V., {et~al.} 2021, \aap,
  646, A77, \dodoi{10.1051/0004-6361/202039765}

\bibitem[{{Hall} {et~al.}(2007){Hall}, {Lockwood}, \& {Skiff}}]{Hall2007}
{Hall}, J.~C., {Lockwood}, G.~W., \& {Skiff}, B.~A. 2007, \aj, 133, 862,
  \dodoi{10.1086/510356}

\bibitem[{{Hall} {et~al.}(2021){Hall}, {Davies}, {van Saders}, {Nielsen},
  {Lund}, {Chaplin}, {Garc{\'\i}a}, {Amard}, {Breimann}, {Khan}, {See}, \&
  {Tayar}}]{Hall2021}
{Hall}, O.~J., {Davies}, G.~R., {van Saders}, J., {et~al.} 2021, Nature
  Astronomy, 5, 707, \dodoi{10.1038/s41550-021-01335-x}

\bibitem[{{Henry} {et~al.}(1996){Henry}, {Soderblom}, {Donahue}, \&
  {Baliunas}}]{Henry1996}
{Henry}, T.~J., {Soderblom}, D.~R., {Donahue}, R.~A., \& {Baliunas}, S.~L.
  1996, \aj, 111, 439, \dodoi{10.1086/117796}

\bibitem[{{Howard} {et~al.}(2013){Howard}, {Sanchis-Ojeda}, {Marcy}, {Johnson},
  {Winn}, {Isaacson}, {Fischer}, {Fulton}, {Sinukoff}, \&
  {Fortney}}]{Howard2013}
{Howard}, A.~W., {Sanchis-Ojeda}, R., {Marcy}, G.~W., {et~al.} 2013, \nat, 503,
  381, \dodoi{10.1038/nature12767}

\bibitem[{{Hsu} {et~al.}(2019){Hsu}, {Ford}, {Ragozzine}, \& {Ashby}}]{Hsu2019}
{Hsu}, D.~C., {Ford}, E.~B., {Ragozzine}, D., \& {Ashby}, K. 2019, \aj, 158,
  109, \dodoi{10.3847/1538-3881/ab31ab}

\bibitem[{{Isaacson} \& {Fischer}(2010)}]{Isaacson2010}
{Isaacson}, H., \& {Fischer}, D. 2010, \apj, 725, 875,
  \dodoi{10.1088/0004-637X/725/1/875}

\bibitem[{{Johnson} {et~al.}(2017){Johnson}, {Petigura}, {Fulton}, {Marcy},
  {Howard}, {Isaacson}, {Hebb}, {Cargile}, {Morton}, {Weiss}, {Winn}, {Rogers},
  {Sinukoff}, \& {Hirsch}}]{CKS2}
{Johnson}, J.~A., {Petigura}, E.~A., {Fulton}, B.~J., {et~al.} 2017, \aj, 154,
  108, \dodoi{10.3847/1538-3881/aa80e7}

\bibitem[{{Judge} \& {Saar}(2007)}]{Judge2007}
{Judge}, P.~G., \& {Saar}, S.~H. 2007, \apj, 663, 643, \dodoi{10.1086/513004}

\bibitem[{{Kolbl} {et~al.}(2015){Kolbl}, {Marcy}, {Isaacson}, \&
  {Howard}}]{Kolbl2015}
{Kolbl}, R., {Marcy}, G.~W., {Isaacson}, H., \& {Howard}, A.~W. 2015, \aj, 149,
  18, \dodoi{10.1088/0004-6256/149/1/18}

\bibitem[{{Kosiarek} \& {Crossfield}(2020)}]{Kosiarek2020}
{Kosiarek}, M.~R., \& {Crossfield}, I. J.~M. 2020, \aj, 159, 271,
  \dodoi{10.3847/1538-3881/ab8d3a}

\bibitem[{{Kraft}(1967)}]{Kraft1967}
{Kraft}, R.~P. 1967, \apj, 150, 551, \dodoi{10.1086/149359}

\bibitem[{{Lissauer} {et~al.}(2014){Lissauer}, {Marcy}, {Bryson}, {Rowe},
  {Jontof-Hutter}, {Agol}, {Borucki}, {Carter}, {Ford}, {Gilliland}, {Kolbl},
  {Star}, {Steffen}, \& {Torres}}]{Lissauer2014}
{Lissauer}, J.~J., {Marcy}, G.~W., {Bryson}, S.~T., {et~al.} 2014, \apj, 784,
  44, \dodoi{10.1088/0004-637X/784/1/44}

\bibitem[{{Lopez} \& {Fortney}(2013)}]{Lopez2013}
{Lopez}, E.~D., \& {Fortney}, J.~J. 2013, \apj, 776, 2,
  \dodoi{10.1088/0004-637X/776/1/2}

\bibitem[{{Lubin} {et~al.}(2021){Lubin}, {Robertson}, {Stefansson}, {Ninan},
  {Mahadevan}, {Endl}, {Ford}, {Wright}, {Beard}, {Bender}, {Cochran},
  {Diddams}, {Fredrick}, {Halverson}, {Kanodia}, {Metcalf}, {Ramsey}, {Roy},
  {Schwab}, \& {Terrien}}]{Lubin2021}
{Lubin}, J., {Robertson}, P., {Stefansson}, G., {et~al.} 2021, \aj, 162, 61,
  \dodoi{10.3847/1538-3881/ac0057}

\bibitem[{{Luhn} {et~al.}(2022){Luhn}, {Wright}, {Henry}, {Saar}, \&
  {Baum}}]{Luhn2022}
{Luhn}, J.~K., {Wright}, J.~T., {Henry}, G.~W., {Saar}, S.~H., \& {Baum}, A.~C.
  2022, \apjl, 936, L23, \dodoi{10.3847/2041-8213/ac8b13}

\bibitem[{{Mamajek} \& {Hillenbrand}(2008)}]{Mamajek2008}
{Mamajek}, E.~E., \& {Hillenbrand}, L.~A. 2008, \apj, 687, 1264,
  \dodoi{10.1086/591785}

\bibitem[{{Masuda}(2022)}]{Masuda2022}
{Masuda}, K. 2022, \apj, 933, 195, \dodoi{10.3847/1538-4357/ac7527}

\bibitem[{{Mayor} \& {Queloz}(1995)}]{Mayor1995}
{Mayor}, M., \& {Queloz}, D. 1995, \nat, 378, 355, \dodoi{10.1038/378355a0}

\bibitem[{{Mazeh} {et~al.}(2015){Mazeh}, {Perets}, {McQuillan}, \&
  {Goldstein}}]{Mazeh2015}
{Mazeh}, T., {Perets}, H.~B., {McQuillan}, A., \& {Goldstein}, E.~S. 2015,
  \apj, 801, 3, \dodoi{10.1088/0004-637X/801/1/3}

\bibitem[{{McQuillan} {et~al.}(2014){McQuillan}, {Mazeh}, \&
  {Aigrain}}]{McQuillan2014}
{McQuillan}, A., {Mazeh}, T., \& {Aigrain}, S. 2014, \apjs, 211, 24,
  \dodoi{10.1088/0067-0049/211/2/24}

\bibitem[{{Metcalfe} \& {Egeland}(2019)}]{Metcalfe2019}
{Metcalfe}, T.~S., \& {Egeland}, R. 2019, \apj, 871, 39,
  \dodoi{10.3847/1538-4357/aaf575}

\bibitem[{{Metcalfe} {et~al.}(2016){Metcalfe}, {Egeland}, \& {van
  Saders}}]{Metcalfe2016}
{Metcalfe}, T.~S., {Egeland}, R., \& {van Saders}, J. 2016, \apjl, 826, L2,
  \dodoi{10.3847/2041-8205/826/1/L2}

\bibitem[{{Mittag} {et~al.}(2013){Mittag}, {Schmitt}, \&
  {Schr{\"o}der}}]{Mittag2013}
{Mittag}, M., {Schmitt}, J.~H.~M.~M., \& {Schr{\"o}der}, K.~P. 2013, \aap, 549,
  A117, \dodoi{10.1051/0004-6361/201219868}

\bibitem[{{Noyes} {et~al.}(1984){Noyes}, {Hartmann}, {Baliunas}, {Duncan}, \&
  {Vaughan}}]{Noyes1984}
{Noyes}, R.~W., {Hartmann}, L.~W., {Baliunas}, S.~L., {Duncan}, D.~K., \&
  {Vaughan}, A.~H. 1984, \apj, 279, 763, \dodoi{10.1086/161945}

\bibitem[{{Owen} \& {Wu}(2013)}]{Owen2013}
{Owen}, J.~E., \& {Wu}, Y. 2013, \apj, 775, 105,
  \dodoi{10.1088/0004-637X/775/2/105}

\bibitem[{{Pepe} {et~al.}(2013){Pepe}, {Cameron}, {Latham}, {Molinari}, {Udry},
  {Bonomo}, {Buchhave}, {Charbonneau}, {Cosentino}, {Dressing}, {Dumusque},
  {Figueira}, {Fiorenzano}, {Gettel}, {Harutyunyan}, {Haywood}, {Horne},
  {Lopez-Morales}, {Lovis}, {Malavolta}, {Mayor}, {Micela}, {Motalebi},
  {Nascimbeni}, {Phillips}, {Piotto}, {Pollacco}, {Queloz}, {Rice}, {Sasselov},
  {S{\'e}gransan}, {Sozzetti}, {Szentgyorgyi}, \& {Watson}}]{Pepe2013}
{Pepe}, F., {Cameron}, A.~C., {Latham}, D.~W., {et~al.} 2013, \nat, 503, 377,
  \dodoi{10.1038/nature12768}

\bibitem[{{Perryman} {et~al.}(1997){Perryman}, {Lindegren}, {Kovalevsky},
  {Hoeg}, {Bastian}, {Bernacca}, {Cr{\'e}z{\'e}}, {Donati}, {Grenon},
  {Grewing}, {van Leeuwen}, {van der Marel}, {Mignard}, {Murray}, {Le Poole},
  {Schrijver}, {Turon}, {Arenou}, {Froeschl{\'e}}, \&
  {Petersen}}]{Perryman1997}
{Perryman}, M.~A.~C., {Lindegren}, L., {Kovalevsky}, J., {et~al.} 1997, \aap,
  323, L49

\bibitem[{{Petigura}(2020)}]{Petigura2020}
{Petigura}, E.~A. 2020, \aj, 160, 89, \dodoi{10.3847/1538-3881/ab9fff}

\bibitem[{{Petigura} {et~al.}(2017){Petigura}, {Howard}, {Marcy}, {Johnson},
  {Isaacson}, {Cargile}, {Hebb}, {Fulton}, {Weiss}, {Morton}, {Winn}, {Rogers},
  {Sinukoff}, {Hirsch}, \& {Crossfield}}]{Petigura2017}
{Petigura}, E.~A., {Howard}, A.~W., {Marcy}, G.~W., {et~al.} 2017, \aj, 154,
  107, \dodoi{10.3847/1538-3881/aa80de}

\bibitem[{{Petigura} {et~al.}(2018){Petigura}, {Marcy}, {Winn}, {Weiss},
  {Fulton}, {Howard}, {Sinukoff}, {Isaacson}, {Morton}, \&
  {Johnson}}]{Petigura2018}
{Petigura}, E.~A., {Marcy}, G.~W., {Winn}, J.~N., {et~al.} 2018, \aj, 155, 89,
  \dodoi{10.3847/1538-3881/aaa54c}

\bibitem[{{Petigura} {et~al.}(2022){Petigura}, {Rogers}, {Isaacson}, {Owen},
  {Kraus}, {Winn}, {MacDougall}, {Howard}, {Fulton}, {Kosiarek}, {Weiss},
  {Behmard}, \& {Blunt}}]{Petigura2022}
{Petigura}, E.~A., {Rogers}, J.~G., {Isaacson}, H., {et~al.} 2022, \aj, 163,
  179, \dodoi{10.3847/1538-3881/ac51e3}

\bibitem[{{Rosenthal} {et~al.}(2021){Rosenthal}, {Fulton}, {Hirsch},
  {Isaacson}, {Howard}, {Dedrick}, {Sherstyuk}, {Blunt}, {Petigura}, {Knutson},
  {Behmard}, {Chontos}, {Crepp}, {Crossfield}, {Dalba}, {Fischer}, {Henry},
  {Kane}, {Kosiarek}, {Marcy}, {Rubenzahl}, {Weiss}, \&
  {Wright}}]{Rosenthal2021}
{Rosenthal}, L.~J., {Fulton}, B.~J., {Hirsch}, L.~A., {et~al.} 2021, \apjs,
  255, 8, \dodoi{10.3847/1538-4365/abe23c}

\bibitem[{{Rowe} {et~al.}(2014){Rowe}, {Bryson}, {Marcy}, {Lissauer},
  {Jontof-Hutter}, {Mullally}, {Gilliland}, {Issacson}, {Ford}, {Howell},
  {Borucki}, {Haas}, {Huber}, {Steffen}, {Thompson}, {Quintana}, {Barclay},
  {Still}, {Fortney}, {Gautier}, {Hunter}, {Caldwell}, {Ciardi}, {Devore},
  {Cochran}, {Jenkins}, {Agol}, {Carter}, \& {Geary}}]{Rowe2014}
{Rowe}, J.~F., {Bryson}, S.~T., {Marcy}, G.~W., {et~al.} 2014, \apj, 784, 45,
  \dodoi{10.1088/0004-637X/784/1/45}

\bibitem[{{Saar}(2006)}]{Saar2006}
{Saar}, S.~H. 2006, in AAS/Solar Physics Division Meeting, Vol.~37, AAS/Solar
  Physics Division Meeting \#37, 12.01

\bibitem[{{Saar}(2011)}]{Saar2011}
{Saar}, S.~H. 2011, in Astronomical Society of the Pacific Conference Series,
  Vol. 448, 16th Cambridge Workshop on Cool Stars, Stellar Systems, and the
  Sun, ed. C.~{Johns-Krull}, M.~K. {Browning}, \& A.~A. {West}, 1239

\bibitem[{{Sanchis-Ojeda} {et~al.}(2014){Sanchis-Ojeda}, {Rappaport}, {Winn},
  {Kotson}, {Levine}, \& {El Mellah}}]{Sanchis-Ojeda2014}
{Sanchis-Ojeda}, R., {Rappaport}, S., {Winn}, J.~N., {et~al.} 2014, \apj, 787,
  47, \dodoi{10.1088/0004-637X/787/1/47}

\bibitem[{{Santos} {et~al.}(2021){Santos}, {Breton}, {Mathur}, \&
  {Garc{\'\i}a}}]{Santos2021}
{Santos}, A.~R.~G., {Breton}, S.~N., {Mathur}, S., \& {Garc{\'\i}a}, R.~A.
  2021, \apjs, 255, 17, \dodoi{10.3847/1538-4365/ac033f}

\bibitem[{{Shah} {et~al.}(2018){Shah}, {Wright}, {Isaacson}, {Howard}, \&
  {Curtis}}]{Shah2018}
{Shah}, S.~P., {Wright}, J.~T., {Isaacson}, H., {Howard}, A.~W., \& {Curtis},
  J.~L. 2018, \apjl, 863, L26, \dodoi{10.3847/2041-8213/aad40c}

\bibitem[{{Skumanich}(1972)}]{Skumanich1972}
{Skumanich}, A. 1972, \apj, 171, 565, \dodoi{10.1086/151310}

\bibitem[{{Su} {et~al.}(2022){Su}, {Zhang}, {Long}, {Han}, {Misra}, {Meng},
  {Pi}, {Yang}, \& {Yang}}]{Su2022}
{Su}, T., {Zhang}, L.-y., {Long}, L., {et~al.} 2022, \apjs, 261, 26,
  \dodoi{10.3847/1538-4365/ac7151}

\bibitem[{{Thompson} {et~al.}(2018){Thompson}, {Coughlin}, {Hoffman},
  {Mullally}, {Christiansen}, {Burke}, {Bryson}, {Batalha}, {Haas},
  {Catanzarite}, {Rowe}, {Barentsen}, {Caldwell}, {Clarke}, {Jenkins}, {Li},
  {Latham}, {Lissauer}, {Mathur}, {Morris}, {Seader}, {Smith}, {Klaus},
  {Twicken}, {Van Cleve}, {Wohler}, {Akeson}, {Ciardi}, {Cochran}, {Henze},
  {Howell}, {Huber}, {Pr{\v{s}}a}, {Ram{\'\i}rez}, {Morton}, {Barclay},
  {Campbell}, {Chaplin}, {Charbonneau}, {Christensen-Dalsgaard}, {Dotson},
  {Doyle}, {Dunham}, {Dupree}, {Ford}, {Geary}, {Girouard}, {Isaacson},
  {Kjeldsen}, {Quintana}, {Ragozzine}, {Shabram}, {Shporer}, {Silva Aguirre},
  {Steffen}, {Still}, {Tenenbaum}, {Welsh}, {Wolfgang}, {Zamudio}, {Koch}, \&
  {Borucki}}]{Thompson2018}
{Thompson}, S.~E., {Coughlin}, J.~L., {Hoffman}, K., {et~al.} 2018, \apjs, 235,
  38, \dodoi{10.3847/1538-4365/aab4f9}

\bibitem[{{Valenti} \& {Fischer}(2005)}]{Valenti2005}
{Valenti}, J.~A., \& {Fischer}, D.~A. 2005, \apjs, 159, 141,
  \dodoi{10.1086/430500}

\bibitem[{{Van Eylen} {et~al.}(2018){Van Eylen}, {Agentoft}, {Lundkvist},
  {Kjeldsen}, {Owen}, {Fulton}, {Petigura}, \& {Snellen}}]{VanEylen2018}
{Van Eylen}, V., {Agentoft}, C., {Lundkvist}, M.~S., {et~al.} 2018, \mnras,
  479, 4786, \dodoi{10.1093/mnras/sty1783}

\bibitem[{{van Saders} {et~al.}(2016){van Saders}, {Ceillier}, {Metcalfe},
  {Silva Aguirre}, {Pinsonneault}, {Garc{\'\i}a}, {Mathur}, \&
  {Davies}}]{vanSaders2016}
{van Saders}, J.~L., {Ceillier}, T., {Metcalfe}, T.~S., {et~al.} 2016, \nat,
  529, 181, \dodoi{10.1038/nature16168}

\bibitem[{{Vaughan} {et~al.}(1978){Vaughan}, {Preston}, \&
  {Wilson}}]{Vaughan1978}
{Vaughan}, A.~H., {Preston}, G.~W., \& {Wilson}, O.~C. 1978, \pasp, 90, 267,
  \dodoi{10.1086/130324}

\bibitem[{{Walkowicz} \& {Basri}(2013)}]{Walkowicz2013}
{Walkowicz}, L.~M., \& {Basri}, G.~S. 2013, \mnras, 436, 1883,
  \dodoi{10.1093/mnras/stt1700}

\bibitem[{{Weiss} {et~al.}(2018){Weiss}, {Marcy}, {Petigura}, {Fulton},
  {Howard}, {Winn}, {Isaacson}, {Morton}, {Hirsch}, {Sinukoff}, {Cumming},
  {Hebb}, \& {Cargile}}]{Weiss2018}
{Weiss}, L.~M., {Marcy}, G.~W., {Petigura}, E.~A., {et~al.} 2018, \aj, 155, 48,
  \dodoi{10.3847/1538-3881/aa9ff6}

\bibitem[{{Weiss} {et~al.}(2024){Weiss}, {Isaacson}, {Howard}, {Fulton},
  {Petigura}, {Fabrycky}, {Jontof-Hutter}, {Steffen}, {Schlichting}, {Wright},
  {Beard}, {Brinkman}, {Chontos}, {Giacalone}, {Hill}, {Kosiarek},
  {MacDougall}, {Mo{\v{c}}nik}, {Polanski}, {Turtelboom}, {Tyler}, \& {Van
  Zandt}}]{Weiss2024}
{Weiss}, L.~M., {Isaacson}, H., {Howard}, A.~W., {et~al.} 2024, \apjs, 270, 8,
  \dodoi{10.3847/1538-4365/ad0cab}

\bibitem[{{Wright}(2004)}]{Wright2004b}
{Wright}, J.~T. 2004, \aj, 128, 1273, \dodoi{10.1086/423221}

\bibitem[{{Wright} {et~al.}(2004){Wright}, {Marcy}, {Butler}, \&
  {Vogt}}]{Wright2004}
{Wright}, J.~T., {Marcy}, G.~W., {Butler}, R.~P., \& {Vogt}, S.~S. 2004, ApJS,
  152, 261, \dodoi{10.1086/386283}

\bibitem[{{Wright} {et~al.}(2008){Wright}, {Marcy}, {Butler}, {Vogt}, {Henry},
  {Isaacson}, \& {Howard}}]{Wright2008}
{Wright}, J.~T., {Marcy}, G.~W., {Butler}, R.~P., {et~al.} 2008, \apjl, 683,
  L63, \dodoi{10.1086/587461}

\bibitem[{{Zhang} {et~al.}(2020){Zhang}, {Bi}, {Li}, {Jiang}, {Li}, {He}, {Yu},
  {Khanna}, {Ge}, {Liu}, {Tian}, {Wu}, \& {Zhang}}]{Zhang2020}
{Zhang}, J., {Bi}, S., {Li}, Y., {et~al.} 2020, \apjs, 247, 9,
  \dodoi{10.3847/1538-4365/ab6165}

\end{thebibliography}
\bibliographystyle{aasjournal}

\end{document}